\documentclass[]{aastex631}

\shorttitle{Effect of Velocity on Kilonova}
\shortauthors{Fryer et al.}
\graphicspath{{./}{figures/}}
\def\RIT{Center for Computational Relativity and Gravitation, Rochester Institute of Technology, Rochester, New York 14623, USA}

\begin{document}

\title{The Effect of the Velocity Distribution on Kilonova Emission}

\correspondingauthor{Chris L. Fryer}
\email{fryer@lanl.gov}

\author[0000-0003-2624-0056]{Chris L. Fryer}
\affiliation{Center for Theoretical Astrophysics, Los Alamos National Laboratory, Los Alamos, NM, 87545, USA}
\affiliation{Computer, Computational, and Statistical Sciences Division, Los Alamos National Laboratory, Los Alamos, NM, 87545, USA}
\affiliation{The University of Arizona, Tucson, AZ 85721, USA}
\affiliation{Department of Physics and Astronomy, The University of New Mexico, Albuquerque, NM 87131, USA}
\affiliation{The George Washington University, Washington, DC 20052, USA}
\author[0000-0001-6893-0608]{Aimee~L. Hungerford}
\affiliation{Center for Theoretical Astrophysics, Los Alamos National Laboratory, Los Alamos, NM, 87545, USA}
\affiliation{Joint Institute for Nuclear Astrophysics - Center for the Evolution of the Elements, USA}
\affiliation{Computational Physics Division, Los Alamos National Laboratory, Los Alamos, NM, 87545, USA}
\author[0000-0003-3265-4079]{Ryan~T. Wollaeger}
\affiliation{Center for Theoretical Astrophysics, Los Alamos National Laboratory, Los Alamos, NM, 87545, USA}
\affiliation{Computer, Computational, and Statistical Sciences Division, Los Alamos National Laboratory, Los Alamos, NM, 87545, USA}
\author[0000-0001-6432-7860]{Jonah M. Miller}
\affiliation{Center for Theoretical Astrophysics, Los Alamos National Laboratory, Los Alamos, NM, 87545, USA}
\affiliation{Computer, Computational, and Statistical Sciences Division, Los Alamos National Laboratory, Los Alamos, NM, 87545, USA}
\author[0000-0002-3316-5149]{Soumi De}
\affiliation{Center for Theoretical Astrophysics, Los Alamos National Laboratory, Los Alamos, NM, 87545, USA}
\affiliation{Computer, Computational, and Statistical Sciences Division, Los Alamos National Laboratory, Los Alamos, NM, 87545, USA}
\author[0000-0003-1087-2964]{Christopher J. Fontes}
\affiliation{Center for Theoretical Astrophysics, Los Alamos National Laboratory, Los Alamos, NM, 87545, USA}
\affiliation{Computational Physics Division, Los Alamos National Laboratory, Los Alamos, NM, 87545, USA}
\author[0000-0003-4156-5342]{Oleg Korobkin}
\affiliation{Center for Theoretical Astrophysics, Los Alamos National Laboratory, Los Alamos, NM, 87545, USA}
\affiliation{Theory Division, Los Alamos National Laboratory, Los Alamos, NM, 87545, USA}
\author[0000-0002-3023-0371]{Atul Kedia}
\affiliation{\RIT}
\author[0000-0001-7042-4472]{Marko Ristic}
\affiliation{\RIT}
\author[0000-0001-5832-8517]{Richard O'Shaughnessy}
\affiliation{\RIT}



\begin{abstract}

The electromagnetic emission from the non-relativistic ejecta launched
in neutron star mergers (either dynamically or through a disk wind)
has the potential to probe both the total mass and composition of this
ejecta.  These observations are crucial in understanding the role of
these mergers in the production of r-process elements in the universe.
However, many properties of the ejecta can alter the light-curves and
we must both identify which properties play a role in shaping this
emission and understand the effects these properties have on the
emission before we can use observations to place strong constraints on
the amount of r-process elements produced in the merger.  This paper
focuses on understanding the effect of the velocity distribution
(amount of mass moving at different velocities) for lanthanide-rich
ejecta on the light-curves and spectra.  The simulations use
distributions guided by recent calculations of disk outflows and
compare the velocity-distribution effects to those of ejecta mass,
velocity and composition.  Our comparisons show that uncertainties in
the velocity distribution can lead to factor of 2-4 uncertainties in
the inferred ejecta mass based on peak infra-red luminosities.  We
also show that early-time UV or optical observations may be able to
constrain the velocity distribution, reducing the uncertainty in the
ejecta mass. 

\end{abstract}

\keywords{ultraviolet astronomy, neutron stars}


\section{Introduction} \label{sec:intro}

Neutron star mergers have long been predicted to be both progenitors of short-duration gamma-ray bursts~\citep{1991AcA....41..257P,1999ApJ...518..356P,1999ApJ...526..152F} and sources for r-process~\citep{1976ApJ...210..549L,1977ApJ...213..225L,1982ApL....22..143S}.  The tidally-ejected material in these mergers is sufficiently neutron-rich to produce a complete r-process signature.  Proving these two theoretical claims has been more difficult.   Gamma-ray burst studies have focused on the claim that, if short-duration bursts were produced by compact mergers, the kicks producing in compact object formation would lead to large proper motions in the binaries, ejecting the systems and predicting offsets of the merger event with respect to star the host-galaxy star-forming~\citep{1999ApJ...526..152F,1999MNRAS.305..763B}.  These offsets were later confirmed by observations~\citep{2010ApJ...708....9F}, strongly supporting the tie between compact mergers and short-duration bursts.  The rates, first derived from pulsar models~\citep{1977ApJ...213..225L} and then through observations of short-duration gamma-ray bursts~\citep{2015ApJ...815..102F}, demonstrated the potential for these mergers to dominate the r-process yields.

The joint detection of gravitational and electromagnetic waves from GW170817 proved that mergers can produce the strongly relativistic jets needed to make the population of short-duration gamma-ray bursts~\citep{2017ApJ...848L..12A}.   The concurrent gamma- and gravitational wave signal showed that at least some gamma-rays were produced in the neutron star merger. the gamma-ray luminosity was low and could be explained by a number of sources.  To argue that this merger produced a gamma-ray burst, scientists argued that the gamma rays came from an off-axis jet.  Radio observations of this event were able to prove that a relativistic jet was produced, confirming the off-axis jet (and hence gamma-ray burst) explanation~\citep{2017ApJ...848L..21A}.  This single event provided the currently most-direct demonstration that mergers produce gamma-ray bursts.

UVOIR observations of this merger provide some of the strongest evidence that these mergers also produce r-process.  The light-curves fit the pre-existing light-curve models of r-process rich ejecta~\citep{2017Natur.551...80K}.  However, fits to these light-curves led to a wide range of estimates for the r-process yield~\citep{2018ApJ...855...99C} and, although some of the observed spectral features are indicative of r-process elements~\citep{2017Natur.551...67P,2022ApJ...939....8D}, no detection is so firm to prove r-process production.  Subsequent studies of the physics behind these light-curves have demonstrated the difficulties in determining the exact yields from the existing observations~\citep{2018MNRAS.478.3298W,2018ApJ...852..109T,2020MNRAS.493.4143F,2020ApJ...899...24E,2020MNRAS.496.1369T,2021ApJ...918...44B,2021ApJ...918...10W}.  Simulations of the ejecta both during the merger and the subsequent accretion disk argue that r-process can be produced in throughout the merger~\citep{2019PhRvD.100b3008M,2023ApJ...945L..13C,2023MNRAS.523.2551K}, but the mass and composition of this ejecta remains uncertain.  These uncertainties must be characterized to use existing and future observations to determine the r-process yields from neutron star mergers and determine their role in r-process production in the Galaxy.  

One of the less-studied uncertainties in modeling the ejecta and light curves of kilonovae is the lack of understanding of the velocity distribution of the ejecta.  The velocity distribution when the ejecta reaches a homologous expansion phase can be described as either the velocity as a function of mass coordinate ($v(m)$), or as is often used in light curve codes, mass as a function of velocity ($m(v)$).  This distribution alters the transient light-curve by altering the evolution of the photosphere (if less mass is moving at high velocities, the photosphere is more quickly positioned at the lower-velocity ejecta).   By varying this velocity distribution, large variations can be produced in the light-curves of both type Ia~\citep{2021ApJ...911...96P} and type II~\citep{2016ApJ...820...74D,2017ApJ...850..133D} supernovae.  Indeed, one reason that different progenitors produce different light-curves is that the density profile of a star produces a different velocity distribution of the ejecta.  For most kilonova models, the ejecta velocity distribution is described by simplified models~\citep{2018MNRAS.478.3298W}.  In this paper, we study the effect of the velocity distribution on kilonova light-curves.

\section{Models and Velocity Profiles}
\label{sec:profiles}

For our spectra and light-curve calculations, we use the {\it SuperNu} Monte Carlo method that couples both Implicit Monte Carlo Methods~\citep{1971JCoPh...8..313F} with Discrete Diffusion Monte Carlo~\citep{2012JCoPh.231.6924D} in optically thick regions.  This code has been used extensively in supernova and kilonova light curve and spectra calculations~\citep[see][and references therein]{2021ApJ...918...10W}.

Heating from radioactive decay assumes in-situ energy deposition from electrons/ions and a gray gamma-ray transport implementation for gamma-ray deposition using the electron-fraction dependent opacity prescription from~\cite{1995ApJ...446..766S}.  This gamma-ray transport has been tested against multi-group gamma-ray methods~\citep{2003ApJ...594..390H}, achieving good agreement in the amount and disposition of the energy deposition~\citep[see Figure 13 of][]{2017ApJ...845..168W}.  The gamma-ray opacity is scaled with the electron fraction following the approach of ~\cite{2016ApJ...829..110B}.   We focus on heavy r-process compositions.  Especially with the potential of upcoming ultraviolet (UV) detections with missions like UltraSAT and UVEX and infrared (IR) measurements of potential kilonova candidates emerging from GRBs~\citep{2017ApJ...843L..34K,2019MNRAS.489.2104T,2022Natur.612..228T}, studying heavy r-process ejecta is becoming increasingly important.  As such, we use a base composition that assumes the r-process ($Y_e=0.19$) yields used in \cite{2021ApJ...918...10W}.  The radioactive heating from the elements in this wind ejecta is derived from the results of the WinNet code~\citep{2012ApJ...750L..22W}, along with a decay network to determine the partitioning of energy among the decay products.  We employ the decay product thermalization model of \cite{2016ApJ...829..110B}.

For our thermal photon transport, we use opacities generated with the LANL suite of atomic physics codes~\citep{2015JPhB...48n4014F}~\citep[for details, see][]{2020MNRAS.493.4143F,2022MNRAS.tmp.2583F} in local thermodynamic equilibrium.  These calculations use all of the lanthanide elements, as well as a single actinide element (uranium) and several lighter elements that act as surrogates to represent the fourth- and fifth-row elements. The first four ion stages are considered for each element when calculating the opacities, which are taken from the same database described in \cite{2021ApJ...918...10W}. A list of elements included in this work are shown in Table~\ref{tab:comp}.  The calculations and implementation of these opacities are described in \cite{2020MNRAS.493.4143F,2022MNRAS.tmp.2583F}.  As our focus is on the velocity distribution of the ejecta, we simplify our ejecta using a spherical outflow.  Our base model assumes $0.1\,M_\odot$ of ejecta with a total energy of $4.5 \times 10^{51}$\,erg.  Although we do vary the ejecta energy, mass and composition, most of our studies focus on varying the velocity distribution.

\begin{table*}
\begin{center}
\begin{tabular}{|lc|lc|lc|}
\hline\hline  
Fe & $7.0\times10^{-11}$ & Ce & $0.027$ & Tb & $0.032$ \\
\hline  
Se & $2.5\times10^{-8}$ & Pr & $9.0\times10^{-6}$ & Dy & $0.28$ \\
\hline  
Br & $0.0050$ & Nd & $2.1\times10^{-4}$ & Ho & $0.023$ \\
\hline  
Zr & $3.1\times10^{-7}$ & Pm & $0.0087$ & Er & $0.087$ \\
\hline  
Pd & $3.6\times10^{-7}$ & Sm &  $0.011$ & Tm & $0.046$ \\
\hline  
Te & $0.047$ & Eu & $5.7\times10^{-4}$  & Yb & $0.084$ \\
\hline  
La & $1.2\times10^{-6}$ & Gd & $0.055$ & U & $0.289$ \\
\hline\hline
\end{tabular}
\caption{Surrogate elements used to determine the opacity and their mass fractions in our standard abundance profiles.}
\label{tab:comp}
\end{center}
\end{table*}

Like many transient light-curve codes, {\it SuperNu} assumes that the ejecta are ballistic at the onset of the calculation, producing a homologous outflow with a prescription for the distribution of the amount of mass moving at different velocities.  Unfortunately, most calculations of neutron star mergers do not follow the ejecta to sufficiently late times to produce such ballistic flows and any prescription at this time is approximate~\citep{Neuweiler_2023}.  To incorporate the variation in both the exact properties of the merger (component masses, spins) and uncertainties in the calculations themselves (e.g. inaccuracies of our ballistic assumption), we consider a range of velocity distributions.    

We use 3 different base velocity profiles:  a power-law profile ($m(v) \propto v^{-\alpha}$) used by many groups~\cite[e.g.][]{2019LRR....23....1M}, the wind profile used by \cite{2018MNRAS.478.3298W} and a final suite of simulations using a phenomenological approximation based on our disk wind models.  The power-law is a simple description of the velocity but provides us with a way to better understand the importance of the velocity distribution.  The \cite{2018MNRAS.478.3298W} prescription is based on a simple explosion picture leveraging methods used for thermonuclear supernovae.  Our most accurate approach to understand the velocity distribution from disk ejecta uses the winds produced in recent disk calculations driven by MHD turbulence, such as \cite{2019PhRvD.100b3008M} and \cite{2020ApJ...902...66M} run with the code $\nu$\texttt{bhlight}~\citep{2019ApJS..241...30M}.  These calculations and calculations like it produce ejecta masses anywhere from 15-40\% of the disk mass with electron fractions ranging from 0.15 to 0.5.  With disk masses ranging from $10^{-5}-0.45\, {\rm M_{\odot}}$~\citep{2023PhRvD.107f3028H}, we can expect disk-wind ejecta masses above $0.1 \, {\rm M_{\odot}}$ on top of dynamical ejecta masses that can exceed $0.02 \, {\rm M_{\odot}}$~\citep{2023PhRvD.107f3028H}.  We focus only on the low electron fraction ejecta in this study.

By comparing all 3 formulations of the velocity distribution, we can assess current uncertainties in our understanding of kilonova light-curves and will be the primary focus of this paper.  Of these, only our disk wind models are new to this community and we describe them in more detail here.

\begin{figure}[ht]
\includegraphics[width=6.0in]{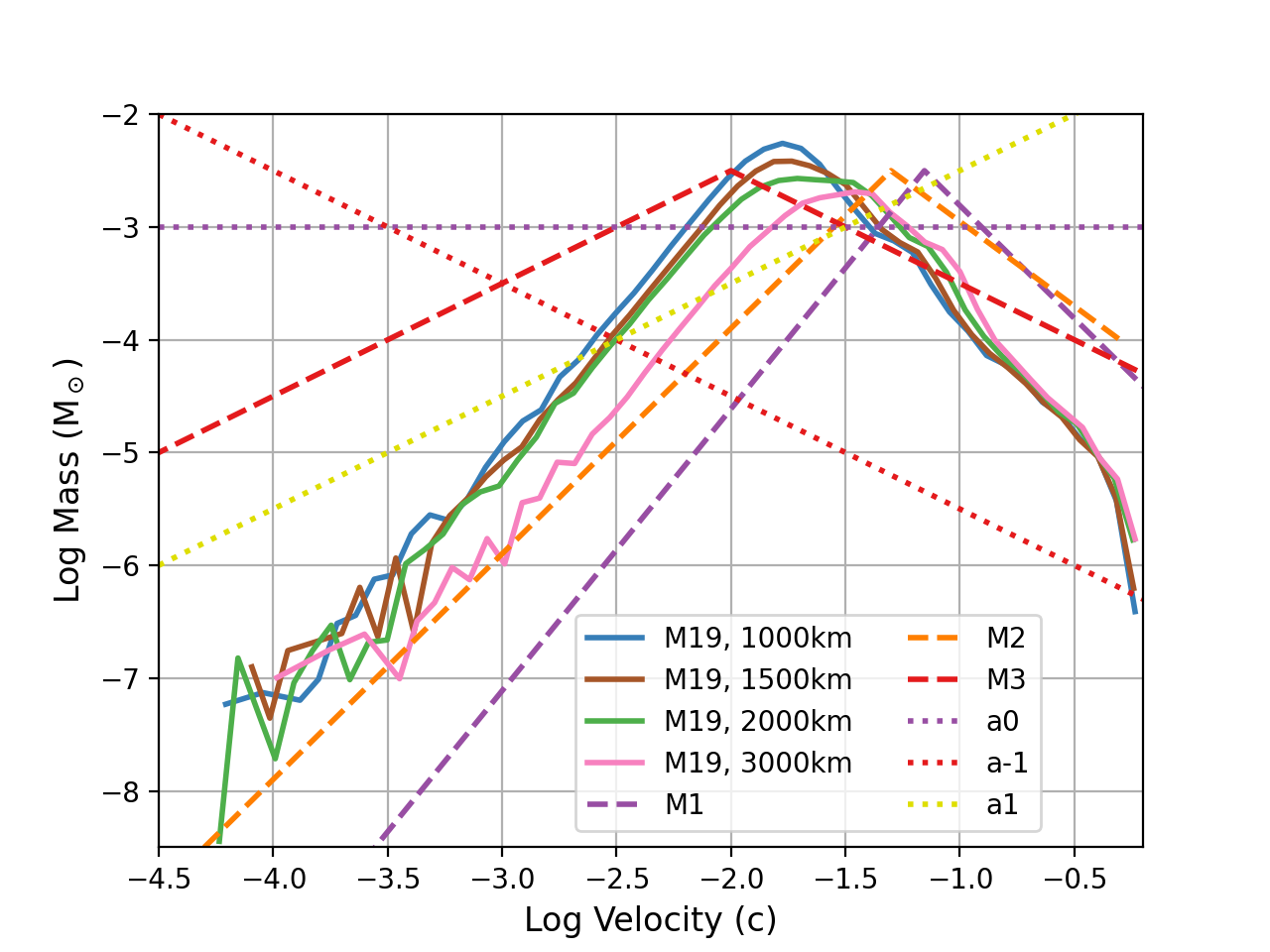}
    \caption{Ejecta mass as a function of velocity for our disk model run out to 1.2s (M19 models using the ~\cite{2019ApJS..241...30M}).  The dashed curves show the M1, M2, M3 2-component power-law models from our study.  Although the shape of these curves is likely to persist in other disk models, we expect a range of solutions and our parameterization is set to cover this range.  The single power-law models are common in the literature.  Although they do not fit the disk models, it is important to include them to understand the differences in the light-curves by those using single power-laws~\cite[e.g.][]{2019LRR....23....1M}.  The ejecta is determined by material moving beyond a critical distance from the merged core using the Bernoulli criteria to determine energy greater than the escape energy.  Even for the late times in these models, the ejecta motion is still evolving with time, producing different results based on different distances.}
    \label{fig:veldisk}
\end{figure}

Figure~\ref{fig:veldisk} shows mass ejected as a function of velocity for one of our disk models. This particular model is designed to match the remnant parameters from numerical relativity simulations of GW170817 \citep{2017PhRvD..96l3012S} focusing on a model that starts after the remnant has collapsed to a black hole. The model begins with a disk of 0.12 $M_\odot$ and electron faction of $Y_e=0.1$ orbiting a black hole of 2.58 $M_\odot$ and dimensionless spin of 0.69. The torus is initially threaded with a single poloidal field loop such that the minimum ratio of gas to magnetic pressure is 100. The simulation was originally run to $\sim$120ms, but it has been extended here to 1.2s. The velocity distribution is determined by measuring tracer particles as they pass through an extraction surface at some large radius, labeled in the figure. Only tracer particles determined to be gravitationally unbound (via both the Bernoulli criteria and velocity greater than escape velocity) are used. 

The velocity distributions of the ejecta in Figure~\ref{fig:veldisk} depend on our choice of the ejection radius.  This demonstrates both the continued evolution of the velocity properties of the ejecta (we are not yet in a homologous outflow) and is also limited by the timescale of our calculation (at 1.2s).  Nevertheless, it is evident that the amount of ejecta mass as a function of velocity $m(v)$ is reasonably well fit by a two-component power-law:
\begin{eqnarray}
    \log{m(v)} &=& \alpha_1 + \beta_1 \log{(v)} {\rm \quad if \, }(v < v_{\rm crit}) \\ 
               &=& \alpha_2 + \beta_2 \log{(v)} {\rm \quad if \, }(v > v_{\rm crit})
\label{eq:diskv}
\end{eqnarray}
where $v$ is the velocity (in units of the speed of light).  We vary $\beta_1, \beta_2$ and $v_{\rm crit}$ to produce different profiles that match the range of our disk wind models.  $\alpha_1$ and $\alpha_2$ are set to produce a continuous function, normalized over the total ejecta mass to match the explosion energy.

\begin{figure}[ht]
\includegraphics[width=6.0in]{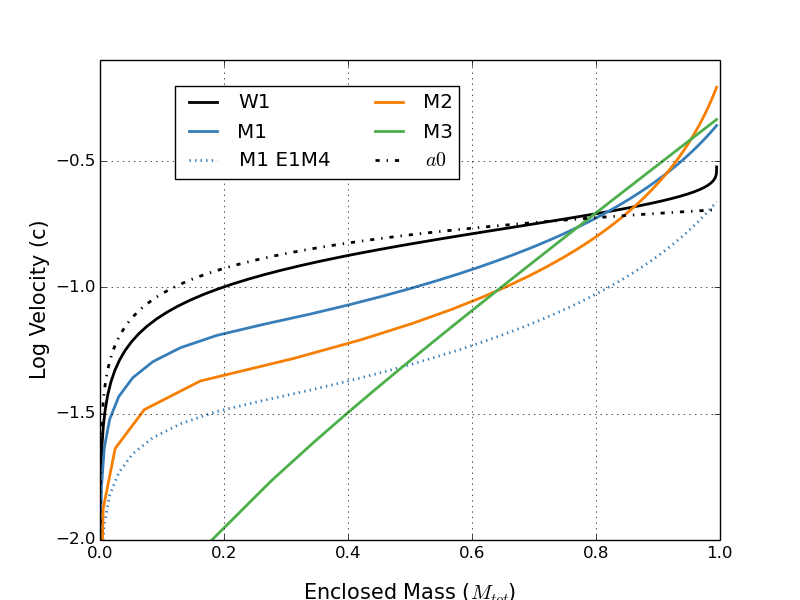}
    \caption{Velocity versus enclosed mass for a set of our models.  All of these models have the same explosion energy and only the M1E1M4 model has a different mass (four times the ejecta mass of the other models).  The fraction of low- versus high-velocity material varies considerably between different models.  As such, the peak luminosity depends on the distribution.}
    \label{fig:veldist}
\end{figure}

The parameters for each of the models used in this study are listed in Table~\ref{tab:models}.  The resultant velocity distributions versus enclosed mass are shown in Figure~\ref{fig:veldist}.  In all of the cases shown in this figure, the total energy is conserved and the masses are all the same.  We do include models that vary the mass and energy.  The velocity distribution remains the same when we scale mass and energy together.  If we alter the mass, but keep the same energy, we scale the velocities in the distribution by a factor to conserve energy.  This alters the position of the $v_{\rm crit}$ although in Table~\ref{tab:models} we provide the $v_{\rm crit}$ prior to normalization for reproducibility.  The broad range of velocity distributions for these models indicates roughly the uncertainties in these velocities.  In this paper, we study how these different distributions affect the observed light-curves.

\begin{table*}
\begin{center}
\begin{tabular}{l|ccccccccc}
\hline\hline              
Model & $E_{\rm Exp}$ & $M_{\rm Exp}$ & $v_{\rm min,max}$ & Velocity Dis. & Composition & $t^{UV}_{peak}$ & $L^{UV}_{peak}$ & $t^{Bol}_{peak}$ & $L^{Bol}_{peak}$  \\
 & ($10^{51}\,{\rm erg}$) & ($M_\odot$) & $10^{7} {\rm cm \, s^{-1}}$, & $v_{\rm crit}, \beta_1, \beta_2$ & & (d) & ($10^{39} {\rm erg \, s^{-1}}$)  & (d) & ($10^{40} {\rm erg \, s^{-1}}$) \\
 & & & $10^{9} {\rm cm \, s^{-1}}$ &  &
& & & \\
\hline
W1 & 4.0 & 0.1 & 14, 9.0 & $W1^{1}$  & Standard & 0.10 & 0.046 & 8.1 & 8.3 \\
W1E2M2 & 8.0 & 0.2 & 14, 9.0 & $W1^{1}$ & Standard & 0.12 & 0.046 & 12.5 & 10.2 \\
W1E4M4 & 16.0 & 0.4 & 14, 9.0 &   $W1^{1}$ & Standard & 0.15 & 0.061 & 18.5 & 12.7 \\
M1 & 4.0 & 0.1 & 7.4, 13 & $0.07c$, $2.5$, $-2$ & Standard & 0.12 & 0.19 & 14.2 & 0.11 \\
M1E.1M.1 & 0.4 & 0.01 & 7.4, 13 & $0.07c$, $2.5$, $-2$ & Standard & 0.04 & 0.016 & 0.09 & 2.2 \\
M1E.1M.2 & 0.4 & 0.02 & 5.2, 9.2 & $0.07c$, $2.5$, $-2$ & Standard & 0.04 & 0.052 & 0.05 & 3.0 \\
M1E.1M.5 & 0.4 & 0.05 & 3.3, 5.8 & $0.07c$, $2.5$, $-2$ & Standard & 0.11 & 0.07 & 14.0 & 2.4 \\
M1E.1M1 & 0.4 & 0.1 & 2.3, 4.1 & $0.07c$, $2.5$, $-2$ & Standard & 0.16 & 0.082 & 0.22 & 4.4 \\
M1E2M2 & 8.0 & 0.2 & 7.4, 13 & $0.07c$, $2.5$, $-2$ & Standard & 0.11 & 0.30 & 0.14 & 21.6 \\
M1E4M4 & 16.0 & 0.4 & 7.4, 13 & $0.07c$, $2.5$, $-2$ & Standard & 0.13 & 0.38 & 0.18 & 32.7 \\
M1E8M8 & 32.0 & 0.8 & 7.4, 13 & $0.07c$, $2.5$, $-2$ & Standard & 0.12 & 1.38 & 48.5 & 0.23 \\
M1E1M2 & 4.0 & 0.2 & 5.2, 9.2 & $0.07c$, $2.5$, $-2$ & Standard & 0.15 & 0.37 & 15.4 & 0.24 \\
M1E1M4 & 4.0 & 0.4 & 3.7, 6.5 & $0.07c$, $2.5$, $-2$ & Standard & 0.19 & 0.40 & 16.6 & 0.43 \\
M1E1M8 & 4.0 & 0.8 & 2.6, 4.6 & $0.07c$, $2.5$, $-2$ & Standard & 0.23 & 0.74 & 19.2 & 0.76 \\
M2 & 4.0 & 0.1 & 10, 19 & $0.05c$, $2.0$, $-1.5$ & Standard & 0.04 & 0.47 & 0.14 & 19.4 \\
M3 & 4.0 & 0.1 & 7.8, 14 & $0.01c$, $1.0$, $-1$ & Standard & 0.12 & 0.24 & 0.04 & 13.4 \\
M1Fe0.01 & 4.0 & 0.1 & 7.4, 13 & $0.07c$, $2.5$, $-2$ & $f_{\rm Fe}=0.01$ & 0.11 & 0.14 & 13.6 & 0.11 \\
M1Fe0.1 & 4.0 & 0.1 & 7.4, 13 & $0.07c$, $2.5$, $-2$ & $f_{\rm Fe}=0.1$ & 0.11 & 0.16 & 13.9 & 0.11 \\
M1Te0.01 & 4.0 & 0.1 & 7.4, 13 & $0.07c$, $2.5$, $-2$ & $f_{\rm Te}=0.01$ & 0.11 & 0.13 & 13.4 & 0.11 \\
M1Te0.5 & 4.0 & 0.1 & 7.4, 13 & $v0.07c$, $2.5$, $-2$ & $f_{\rm Te}=0.5$ & 0.11 & 0.20 & 14.9 & 0.11 \\
M1U0.005 & 4.0 & 0.1 & 7.4, 13 & $0.07c$, $2.5$, $-2$ & $f_{\rm U}=0.005$ & 0.08 & 0.20 & 28.5 & 0.08 \\
M1U0.05 & 4.0 & 0.1 & 7.4, 13 & $0.07c$, $2.5$, $-2$ & $f_{\rm U}=0.05$ & 0.11 & 0.14 & 13.6 & 0.11 \\
M1U0.1 & 4.0 & 0.1 & 7.4, 13 & $0.07c$, $2.5$, $-2$ & $f_{\rm U}=0.1$ & 0.11 & 0.16 & 13.9 & 0.11 \\
M1U0.5 & 4.0 & 0.1 & 7.4, 13 & $0.07c$, $2.5$, $-2$ & $f_{\rm U}=0.5$ & 0.11 & 0.10 & 12.8 & 0.11 \\
M1Zr0.01 & 4.0 & 0.1 & 7.4, 13 & $0.07c$, $2.5$, $-2$ & $f_{\rm Zr}=0.01$ & 0.077 & 0.10 & 8.1 & 8.4 \\
M1Zr0.01 & 4.0 & 0.1 & 7.4, 13 & $0.07c$, $2.5$, $-2$ & $f_{\rm Zr}=0.01$ & 0.077 & 0.10 & 8.1 & 8.4 \\
a-2 & 4.0 & 0.1 & 8.8, 5.6 & $\alpha=-2$ & Standard & 0.80 & 0.035 & 0.93 & 43.4 \\
a-1 & 4.0 & 0.1 & 9.1, 5.8 & $\alpha=-1$ & Standard & 0.72 & 0.32 & 0.83 & 39.9 \\
a0 & 4.0 & 0.1 & 9.6, 6.0 & $\alpha=0$ & Standard & 0.63 & 0.32 & 0.72 & 35.7 \\
a1 & 4.0 & 0.1 & 11, 6.7 & $\alpha=1$ & Standard & 0.51 & 0.28 & 0.58 & 30.7 \\
\hline\hline
\end{tabular}
\caption{Simulation Suite:  $^1$W1=\cite{2018MNRAS.478.3298W} distribution.}
\label{tab:models}
\end{center}
\end{table*}

It is important to note that, although the disk calculations used in this study run out to late times when compared to many past disk models (over 1s), the trajectories are not yet homologous.  Understanding the exact distribution of the outflow remains an active area of research~\citep[for recent results, see][]{2023PhRvD.107b3016N}.  

\section{Light-Curves}
\label{sec:lc}

With our suite of progenitors, and using the opacities and energy generation implemented in our {\it SuperNu} code, we simulate our kilonova models from 1 hour to 30d.  For our light-curves, we focus on four primary wavelength bands:  a bolometric luminosity, a UV-band filter based on the current expected UltraSAT~\citep{2021SPIE11821E..0UA} filter (covering roughly 220--290~nm), a generic v-band luminosity (4500--6500~\AA), and an IR luminosity (1900-25000~\AA).  For the V and IR bands, we use a simple top-hat function for the filter.  Although features do exist in the light-curves that would cause different variations with different filters, these bands are a good subset to show the effects of the different ejecta properties on the light-curve behavior.

\subsection{Dependence on Velocity Distribution}

\begin{figure}[ht]
\includegraphics[width=3.5in]{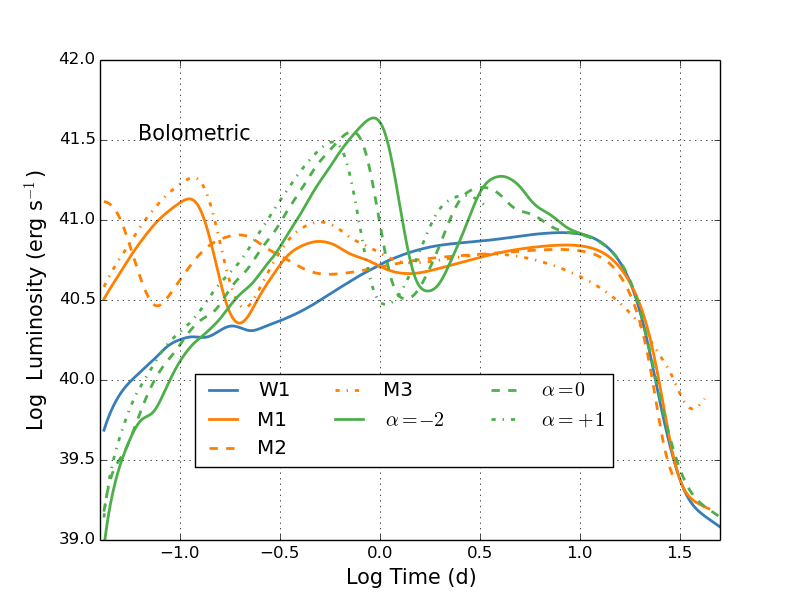}
\includegraphics[width=3.5in]{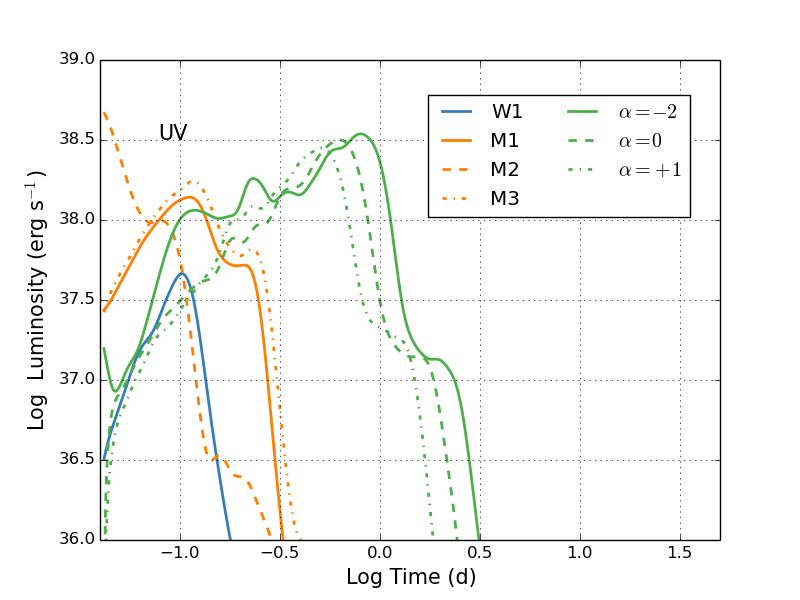}
\includegraphics[width=3.5in]{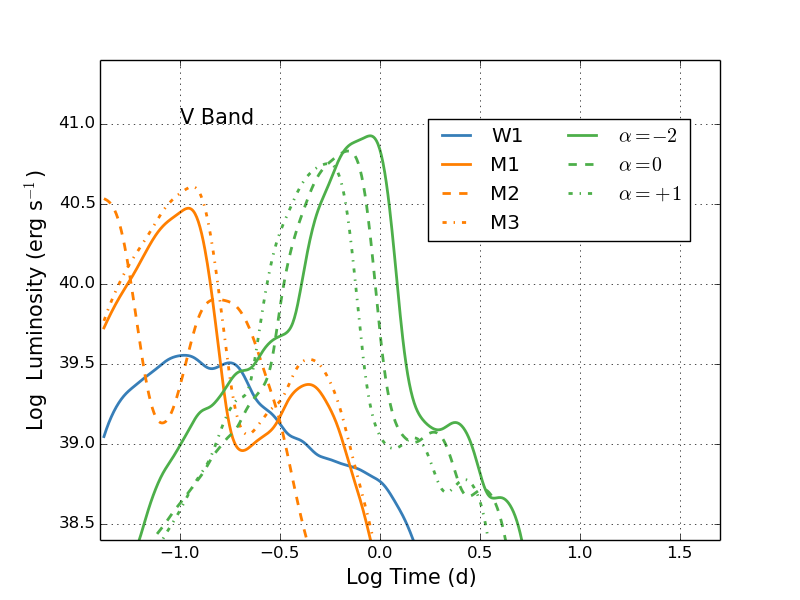}
\includegraphics[width=3.5in]{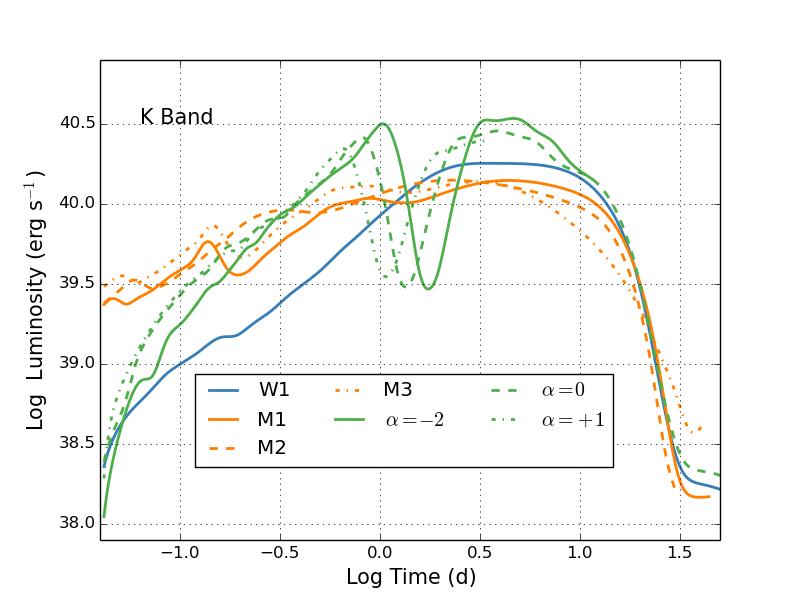}
    \caption{Bolometric, UV, V-band and K-band luminosities for different velocity distributions:  ``W1" corresponds to the standard velocity distribution from \cite{2018MNRAS.478.3298W}, $\alpha$ models correspond to velocity distributions where $m(v) \propto v^{-\alpha}$, M1,M2,M3 correspond to three of our models based on disk models (see Section~\ref{sec:profiles} for more details).  The UV band is based on the UltraSAT~\citep{2021SPIE11821E..0UA} filter covering roughly 220--290~nm, the V-band luminosity assumes a flat-top of 4500--6500~\AA,  and the IR band spas 1900-25000~\AA.  The photosphere moves inward quickly for our $\alpha$ models, allowing them to remain hot longer and peak later.  The photosphere remains further out for more realistic models and the lower temperatures at these photospheres more quickly evolve to longer wavelengths.}
    \label{fig:morph}
\end{figure}

Figure~\ref{fig:morph} shows a subset of the model suite in Table~\ref{tab:models}, focusing on the ejecta velocity distributions including the standard distribution used by many current light-curve calculations (W1), power-law velocity distributions (with varying power-law $\alpha$ values) and our models based on our disk wind models using Equation~\ref{eq:diskv} (for details, see Section~\ref{sec:profiles}).  Especially for the UV and V-bands, the peak emission time can vary by over an order of magnitude ranging from 1~hour to 1~day.  The K-band and bolometric luminosities are much less sensitive to variations in the velocity distribution.  

\begin{figure}
\includegraphics[width=3.5in]{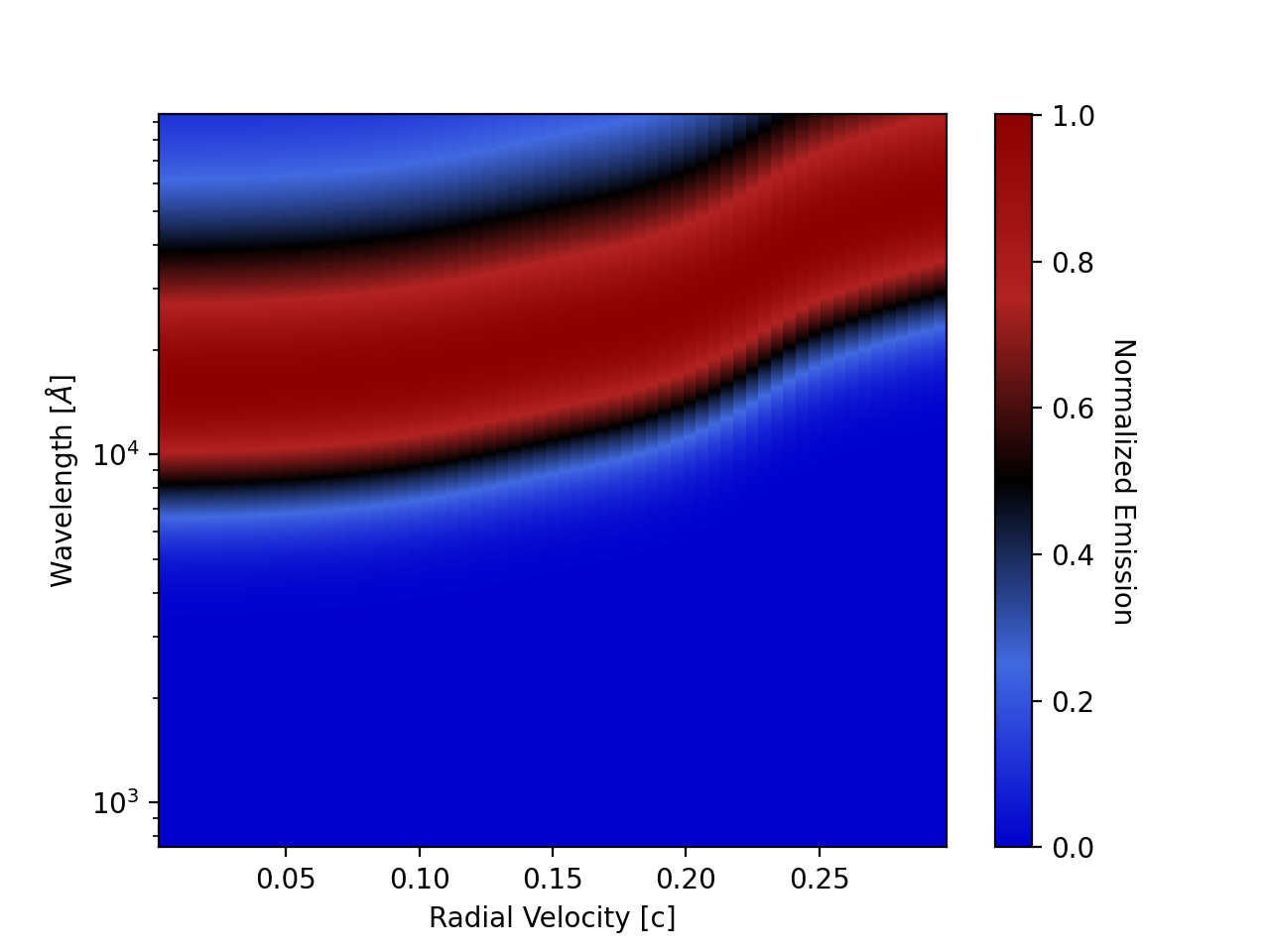}
\includegraphics[width=3.5in]{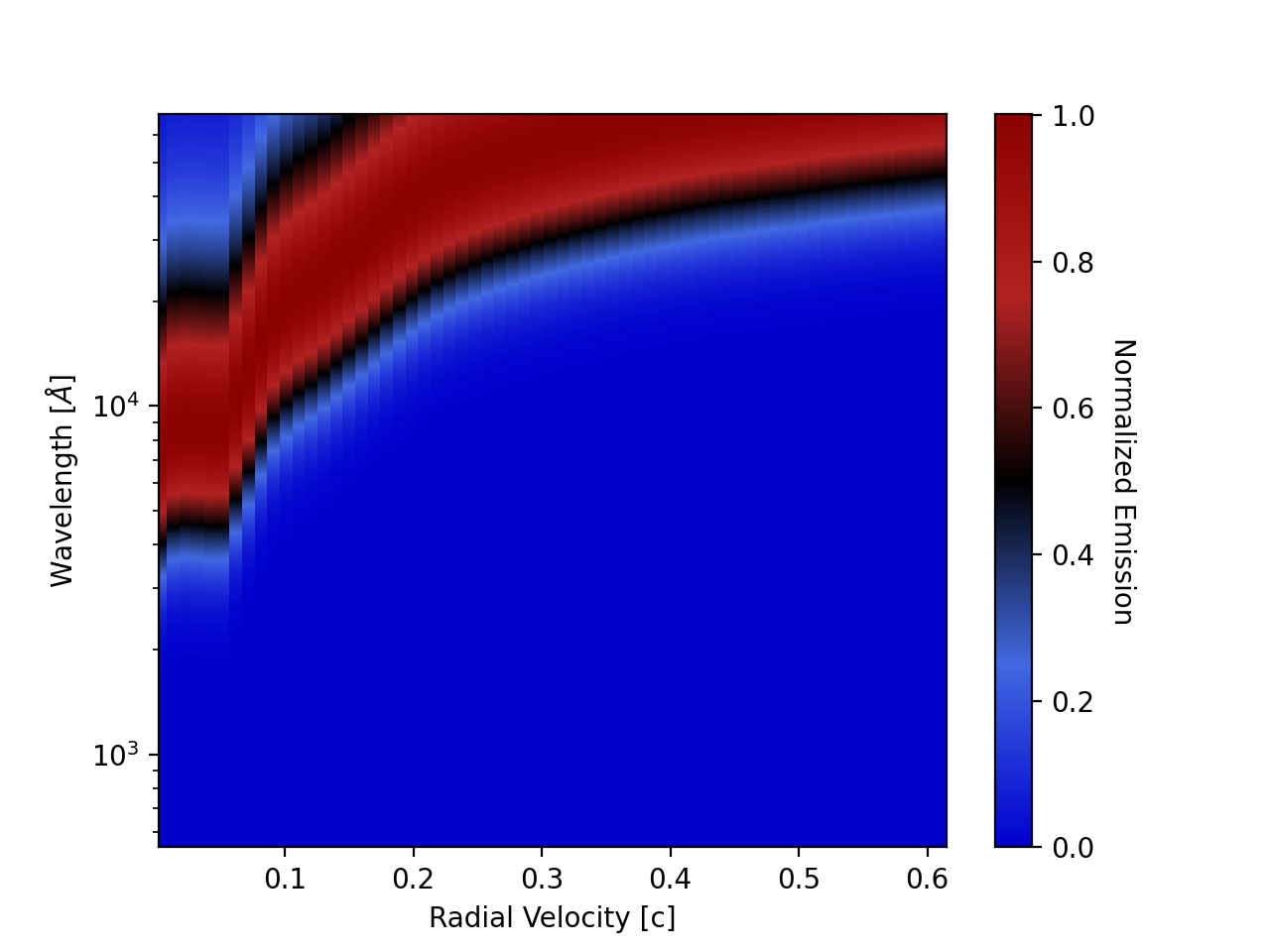}
\includegraphics[width=3.5in]{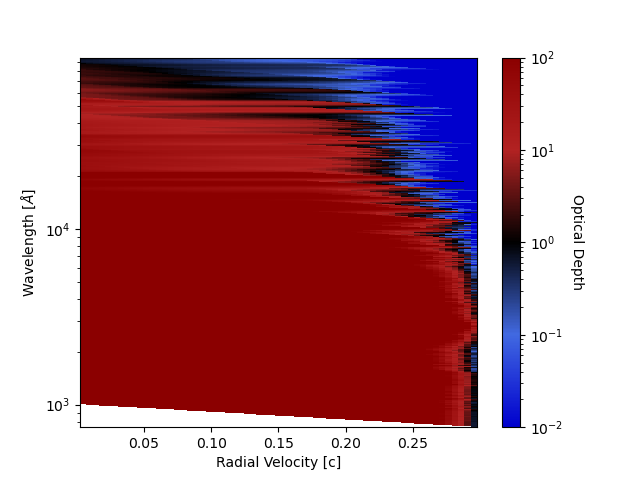}
\includegraphics[width=3.5in]{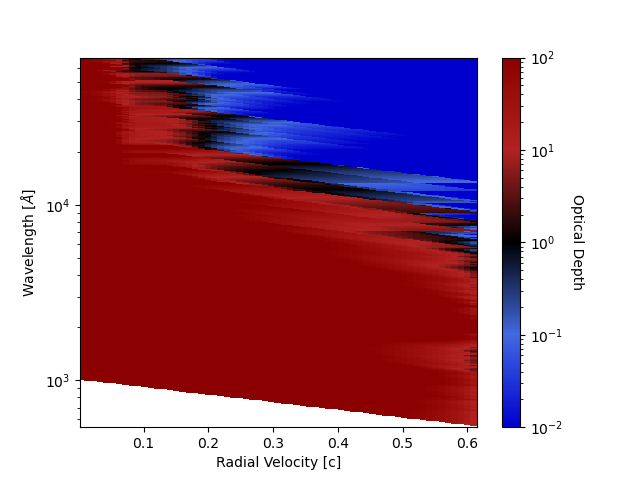}
\includegraphics[width=3.5in]{figclfknstancappla7.97}
\includegraphics[width=3.5in]{figclfknsou3cappla7.97}
    \caption{Normalized Emission (top), optical depth (middle), and an emission weighted optical depth (bottom) at 8\,d as a function of wavelength and velocity space for two models:  W1 (left), M1 (right).  The emission is normalized such that the peak emission in a zone is set to 1.  The emission weighted optical depth focuses on the opacity where the emission is high (if emission is low, the image is transparent/white). In a homologous expansion, the velocity coordinate corresponds to the the radial position.  The emission is calculated assuming a scaled Planckian to determine the location and wavelength range of the emission, the optical depth is calculated using the SuperNu interpolated opacities from the LANL database and the combined plot includes both factors to determine the emission.  The white regions correspond to regions where the emission is negligible.  Deep in the ejecta (low velocities), less expansion has occurred and the material is hotter, producing higher energy (lower wavelength) emission.  But the observed emission depends on opacity.  The higher opacities at high energies tend to limit the high energy emission.  The emission weighted opacity provides a picture of how these effects combine to produce the emerging spectra.}
    \label{fig:suite}
\end{figure}

The wide variation in the light-curves occurs because of differences in the evolution in the photosphere and the characteristics of the ejecta at this photosphere.  One way to understand this evolution is through plots of the photosphere (which varies with wavelength and time) combined with the emission at different radii.  Figure~\ref{fig:suite} shows a series of plots depicting the emission, opacity, and the emission corrected for attenuation, all as a function of both wavelength and velocity (equivalent to radius in a homologous outflow) for two different velocity distributions (W1 and M1) at 8\,d.  If the opacity were low, the observed spectrum would be the integration of the emission across velocity space.  From the top panel of Figure~\ref{fig:suite}, if we could neglect attenuation, we'd expect emission at wavelengths as low as 5,000\AA~ from our kilonovae.  However, the innermost (lowest-velocity) ejecta remains dense at 8\,d and the optical depth is still large.  In the W1 models commonly used in the literature, the optical depth remains high, even at wavelengths above 20,000\AA.  At 8\,d, the escaping radiation is primarily at wavelengths above 20,000\AA.  For our M1 model, the optical depth becomes low even for radiation at 10,000\AA~ and this model will emit at these lower wavelength (higher energy) range.  By combining the emitted spectrum with the optical depth (bottom panels), we produce a picture of the total emission source for these models.  The drastic differences in these plots may be critical in explaining differences between different kilonova light-curve calculations.  Many groups do not describe their velocity distributions and it is clear from these images that the distribution is important.

\begin{figure}
\includegraphics[width=3.5in]{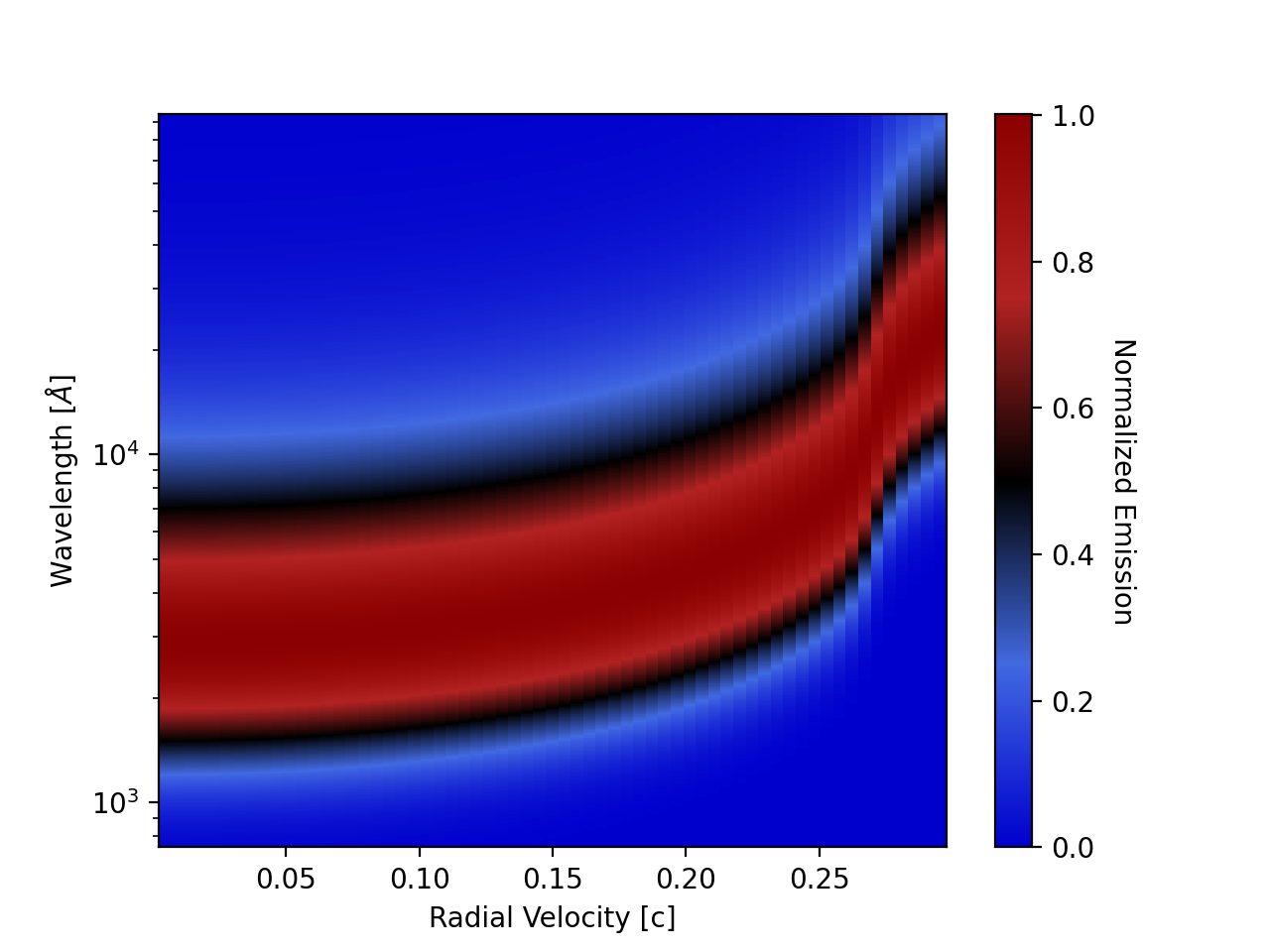}
\includegraphics[width=3.5in]{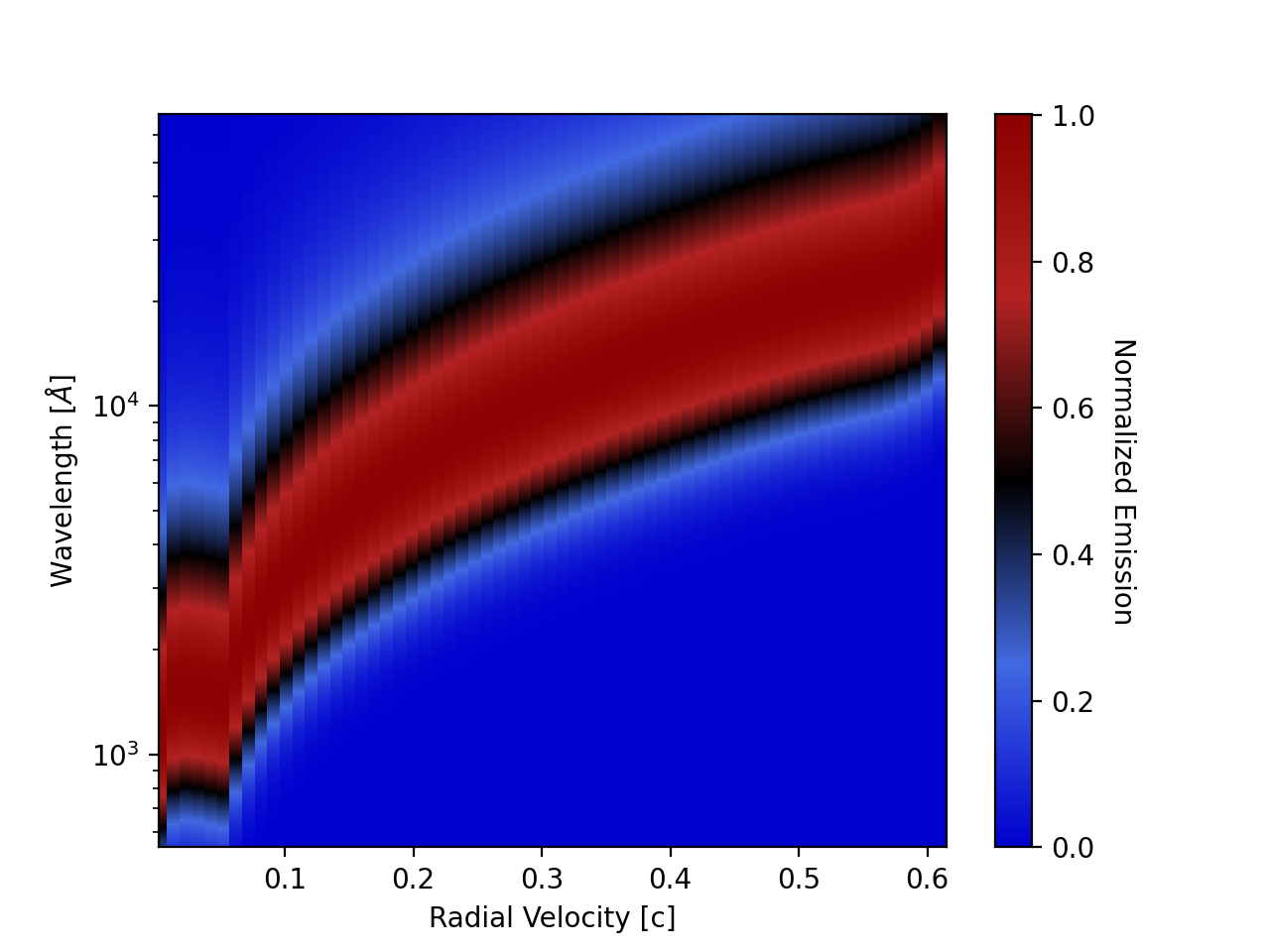}
\includegraphics[width=3.5in]{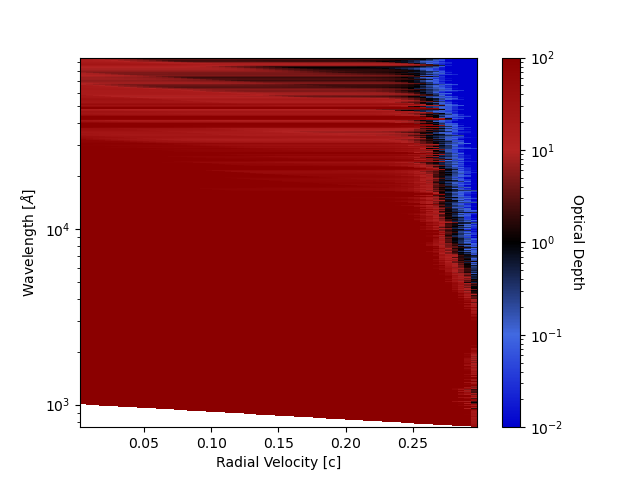}
\includegraphics[width=3.5in]{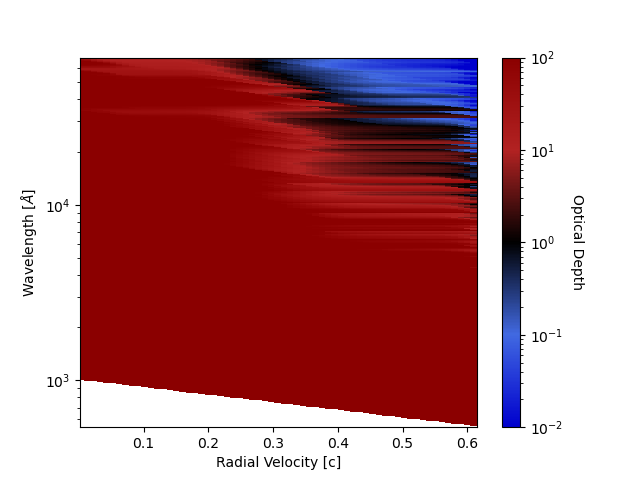}
\includegraphics[width=3.5in]{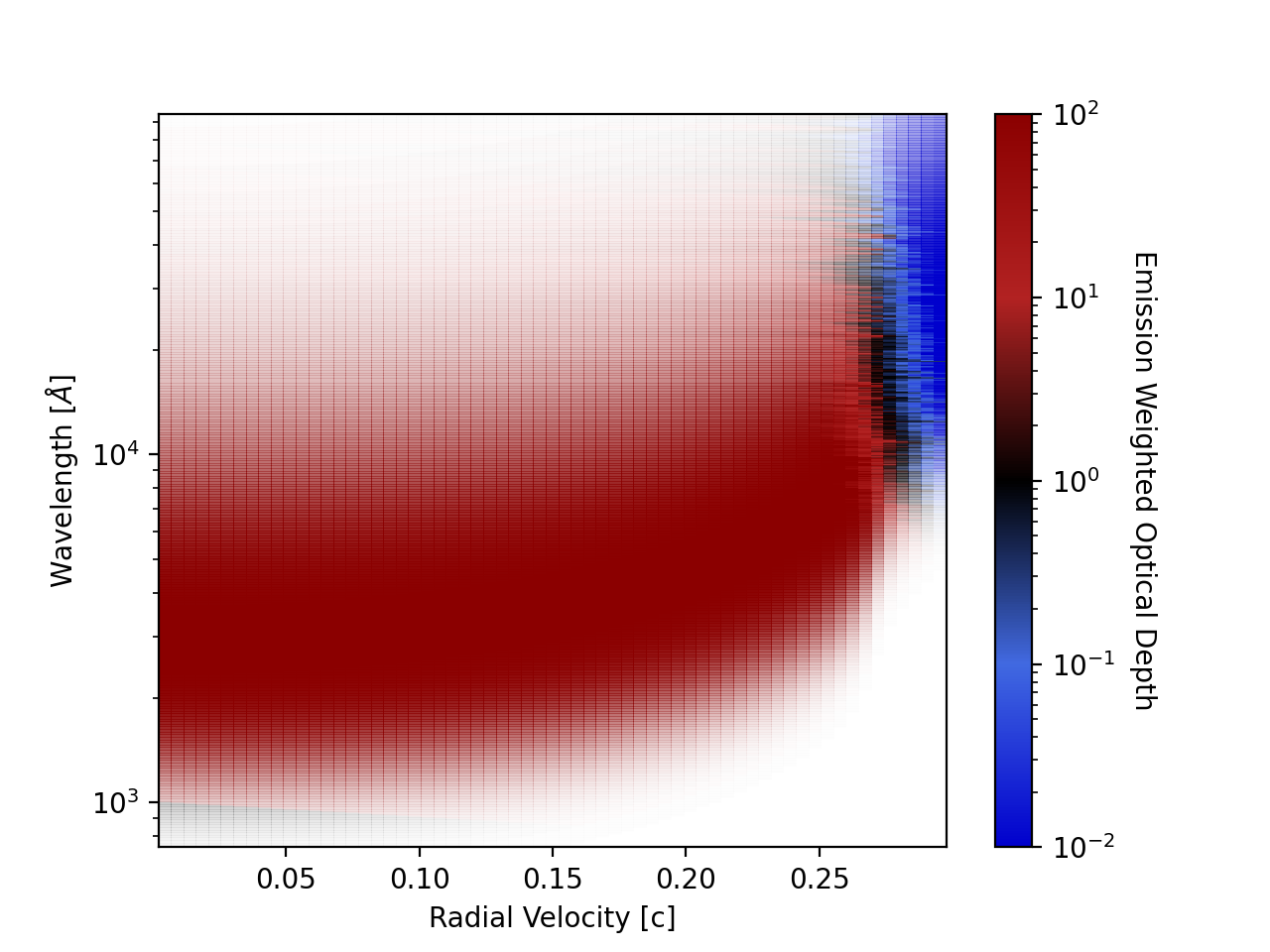}
\includegraphics[width=3.5in]{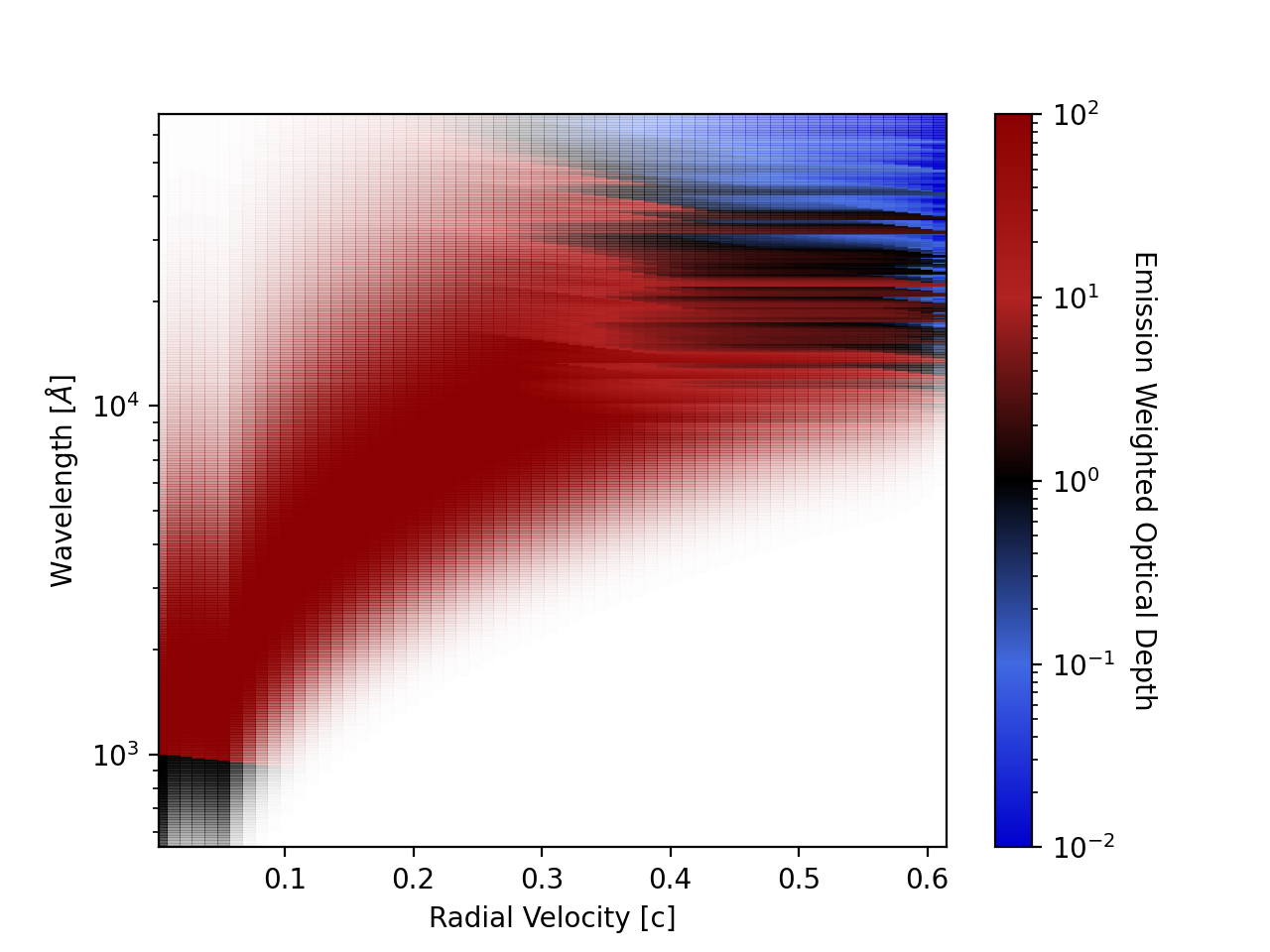}
    \caption{The same set of figures as Figure~\ref{fig:suite} at an earlier epoch (1d).  }
    \label{fig:suite2}
\end{figure}

\begin{figure}
\includegraphics[width=3.5in]{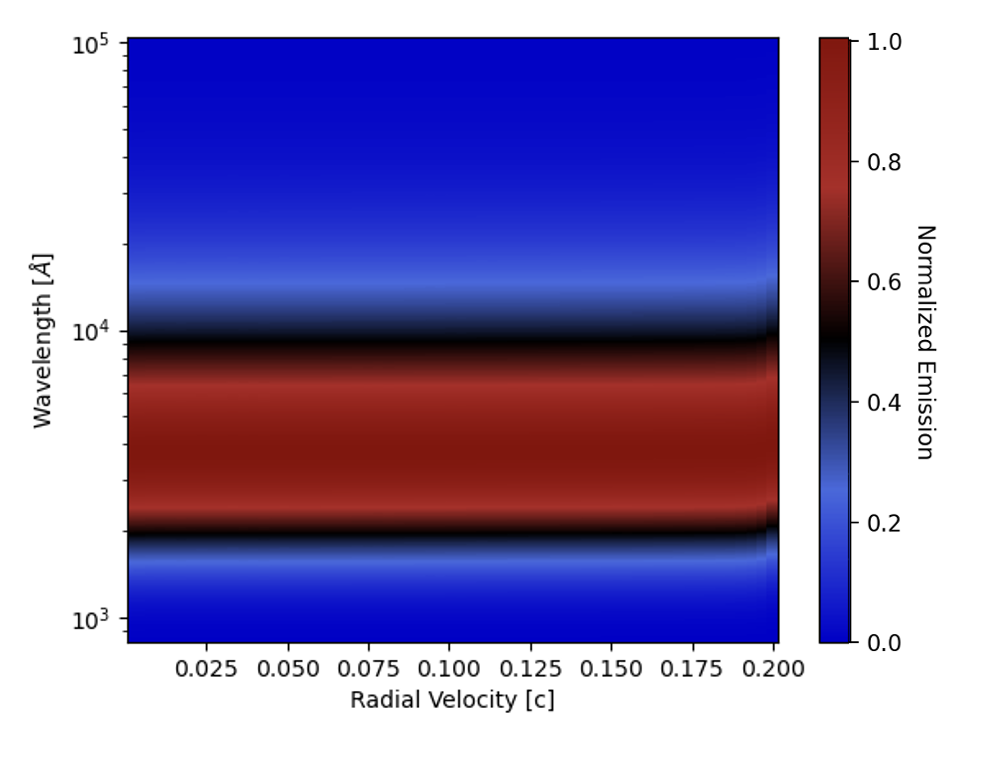}
\includegraphics[width=3.5in]{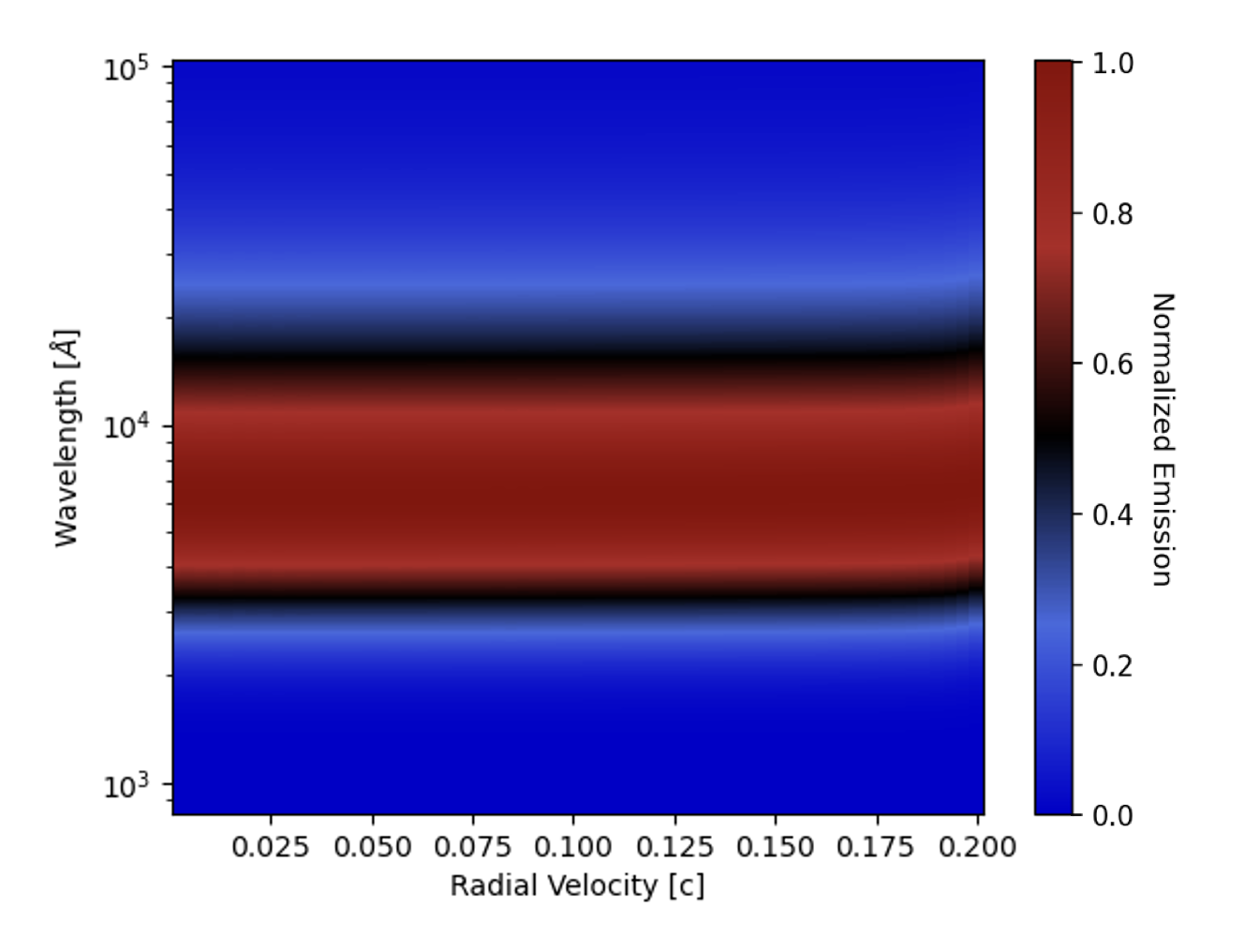}
\includegraphics[width=3.5in]{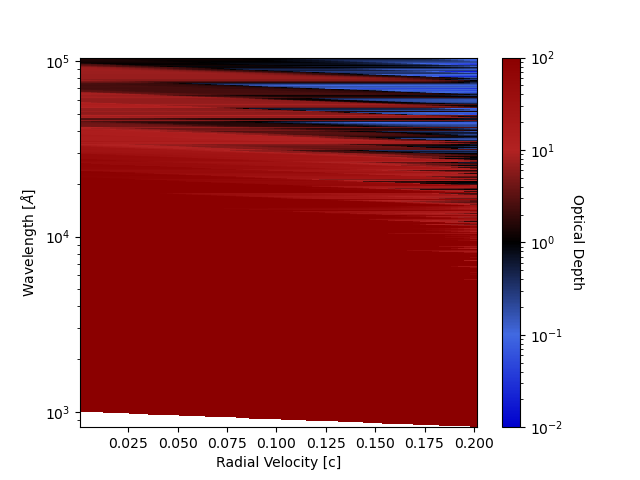}
\includegraphics[width=3.5in]{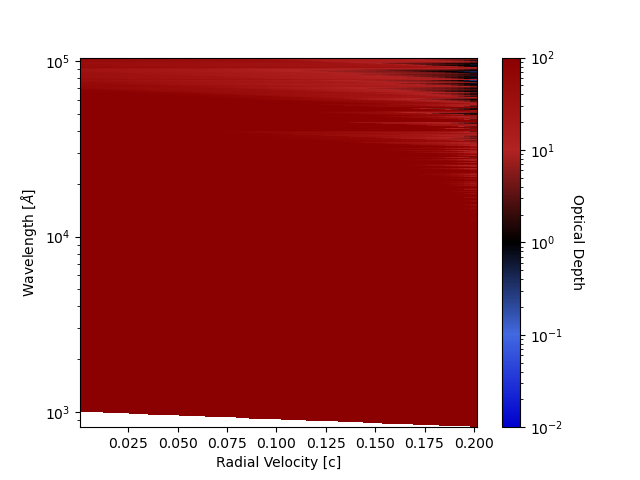}
\includegraphics[width=3.5in]{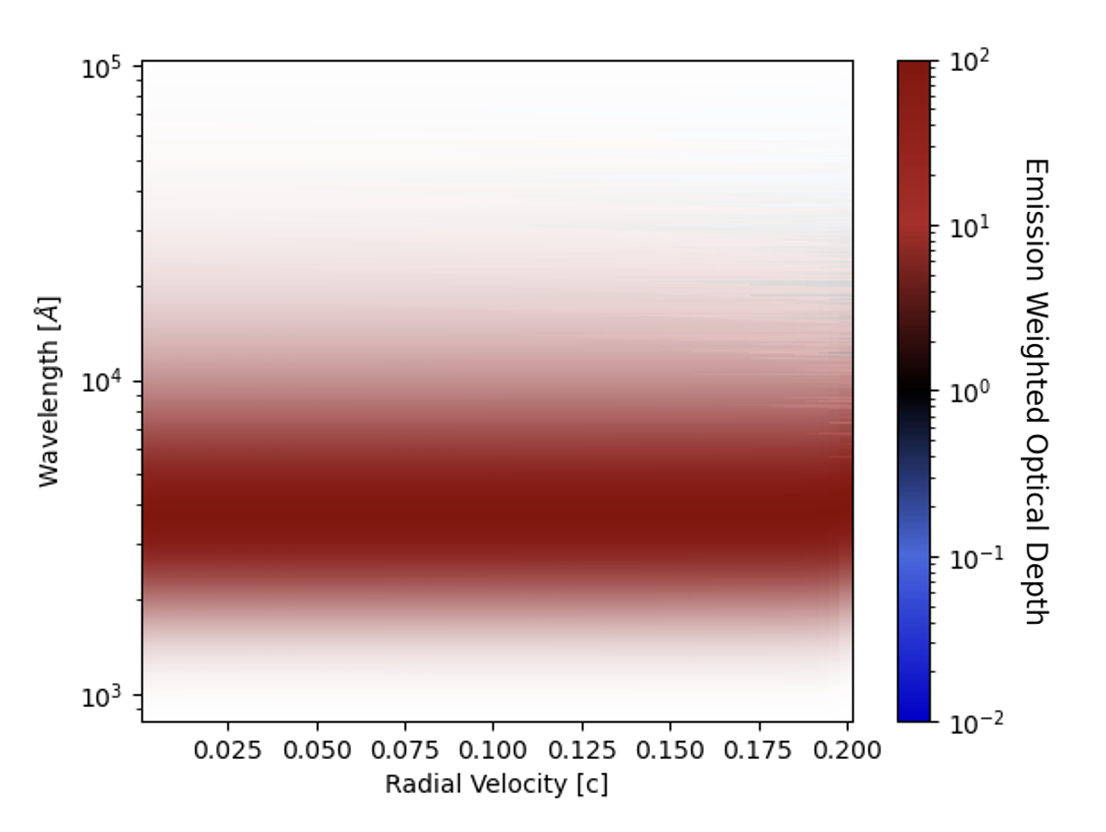}
\includegraphics[width=3.5in]{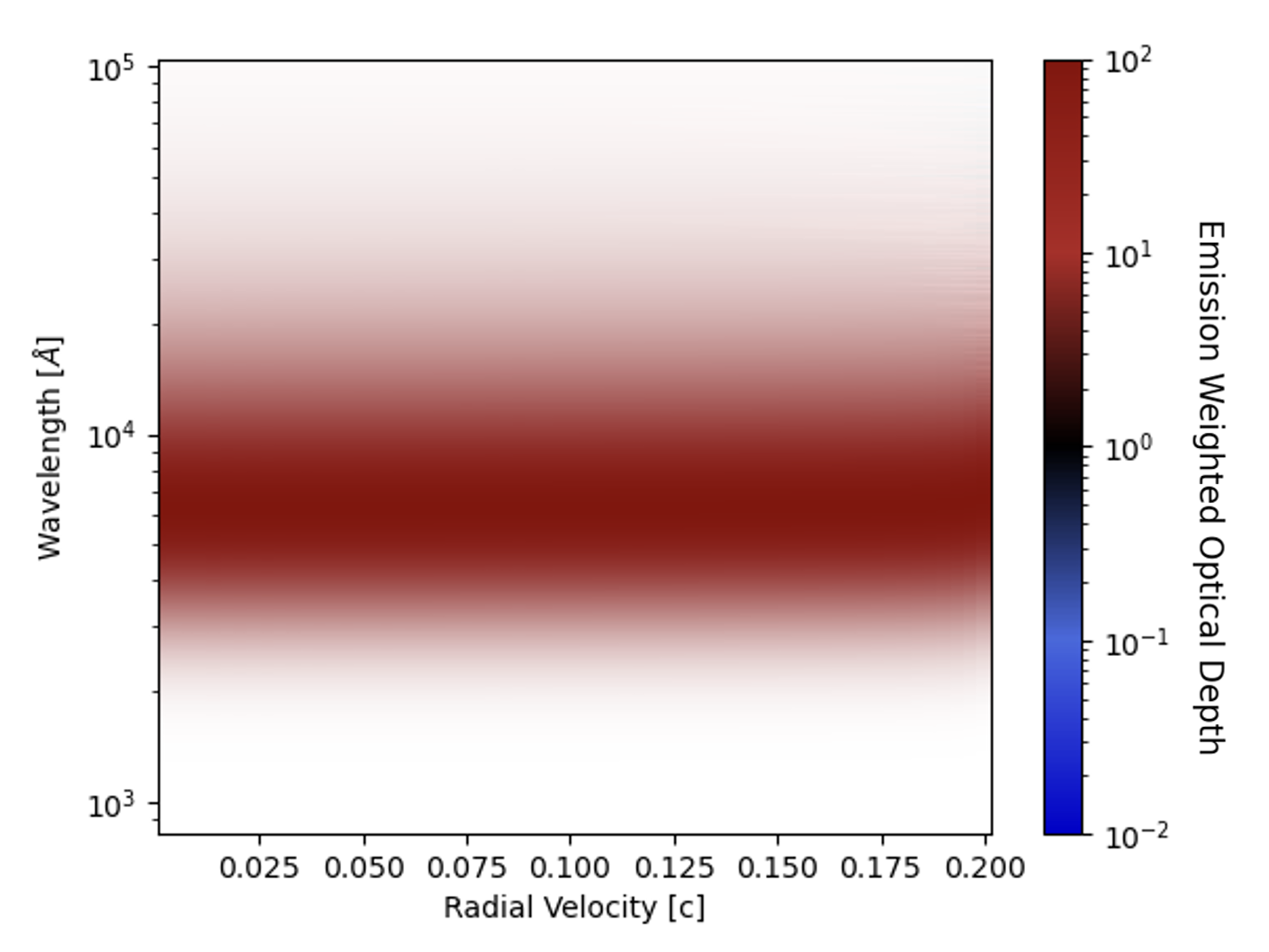}
    \caption{The same set of figures as Figure~\ref{fig:suite} for the $a0$ model at 1 and 2\,d.}
    \label{fig:suitea}
\end{figure}

Figure~\ref{fig:suite2} shows the same information as Figure~\ref{fig:suite}, but at 1\,d instead of 8\,d.  At these early times, the outermost ejecta has expanded and sufficiently cooled so that its emission is primarily in the infra-red.  The innermost ejecta is still emitting in the ultraviolet, but this emission is trapped in the flow.  Figure~\ref{fig:suitea} shows the same information, but for the $a0$ ($\alpha=0$) power-law velocity distribution at 1 and 2\,d.  For most of our models, only a small fraction of material is moving at the highest velocities.  For the $a0$ model, the velocity distribution with mass is constant.  The fastest ejecta in the $a0$ model is much slower than the fastest ejecta in our W1 or two-component (M-series) models.  The slower velocity means that the ejecta remains hotter longer, producing V-band and UV-band emission at later times.  Because this ejecta does not have the mass moving at high-velocities like the W1 model, it does not produce the burst of UV emission at early times.  Note that the peak velocity (x-axis) varies for the different models.  These differences also show just how sensitive the observed emission is to these distributions.

The combined opacity corrected emission plots (bottom panels from Figures~\ref{fig:suite},\ref{fig:suite2},\ref{fig:suitea}) provide a detailed window in understanding the emission arising from a kilonova ejecta.  The velocity of the matter determines both the density and temperature (cooling through adiabatic expansion) of the ejecta.  This determines the emission spectrum but also determines the opacity.  As the temperature decreases, the opacity can actually increase, moving the photosphere outward in mass coordinate.  The observed emission is determined by the conditions at the photosphere.  Shifts in the opacity can alter the position of the photosphere and, hence, the emission can be very sensitive to the velocity distribution of the matter.  The complexity of these models is evident by comparing the conditions of these different models.  Our M-series models tend to have more ejecta at high velocities than the W1 model.  The higher velocity material becomes optically thin more quickly.

\begin{figure}
\includegraphics[width=3.5in]{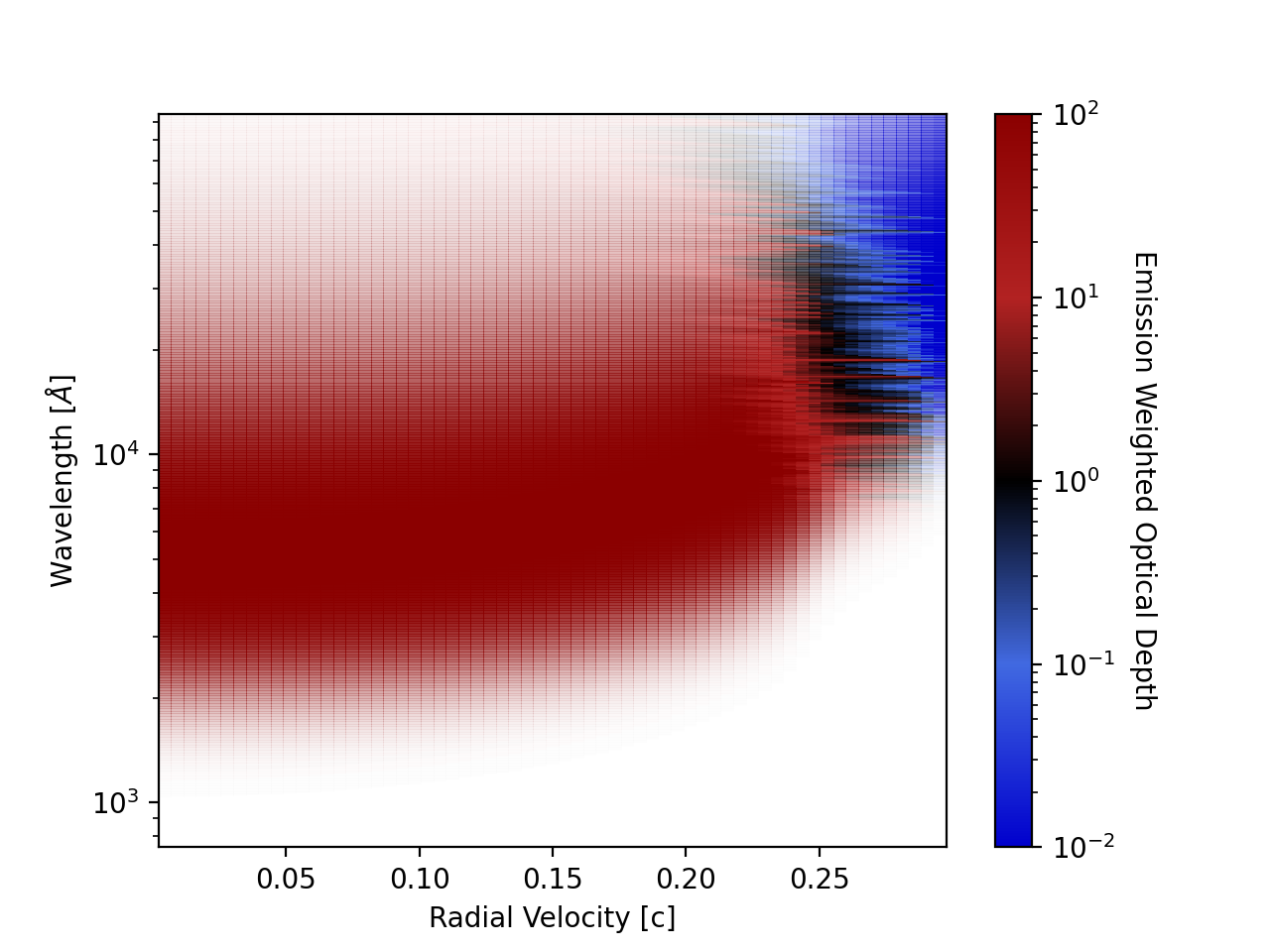}
\includegraphics[width=3.5in]{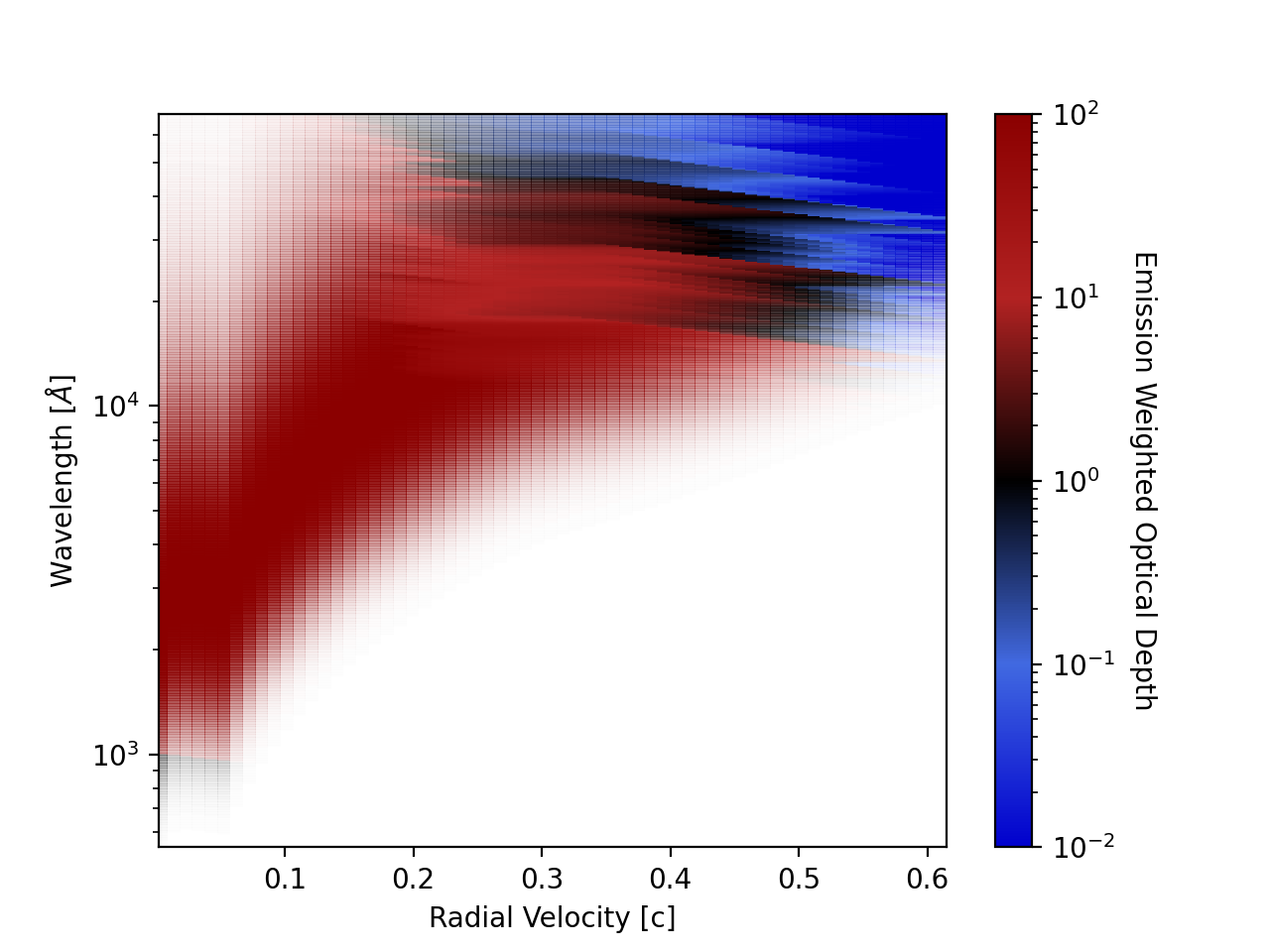}
\includegraphics[width=3.5in]{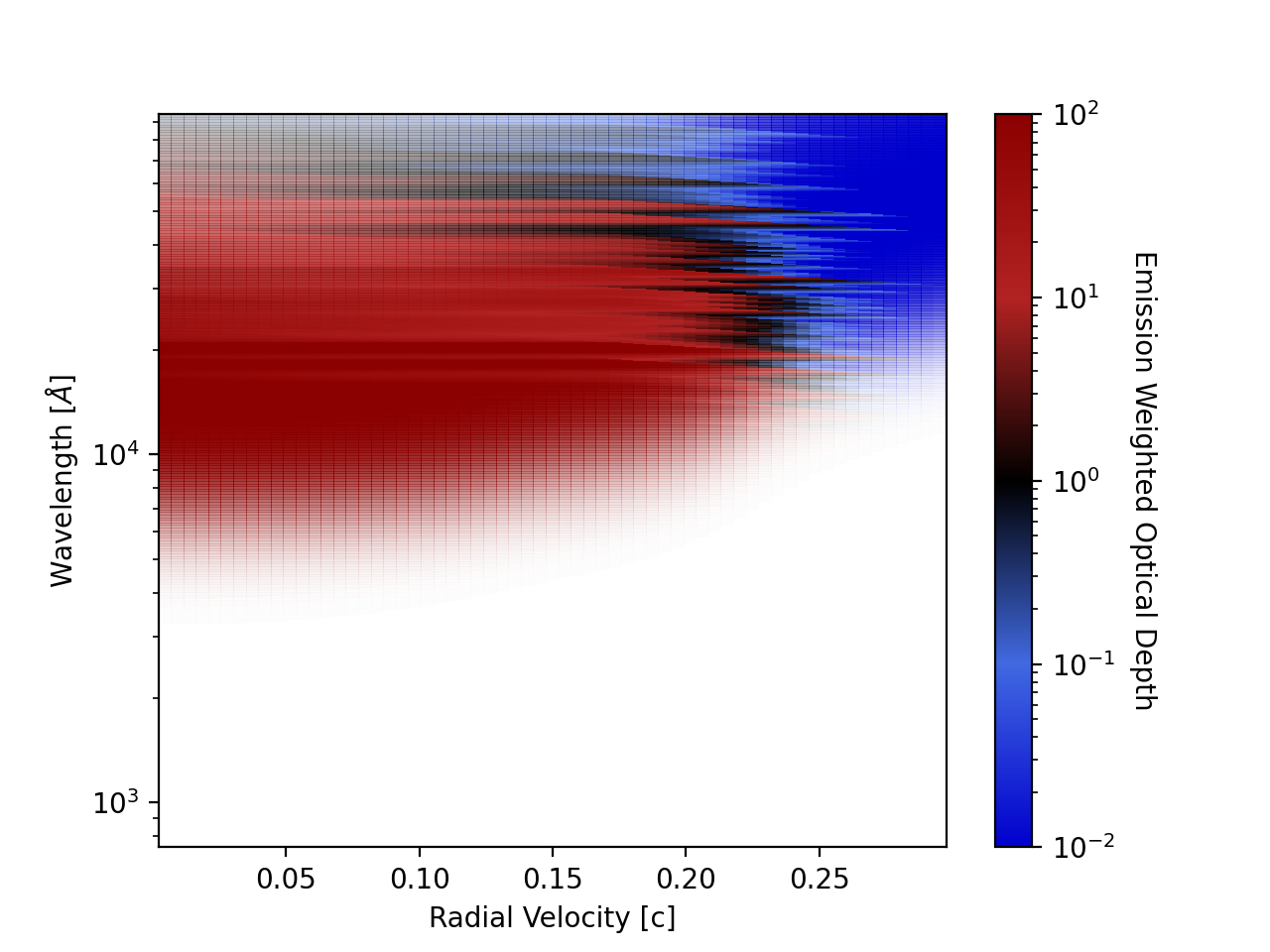}
\includegraphics[width=3.5in]{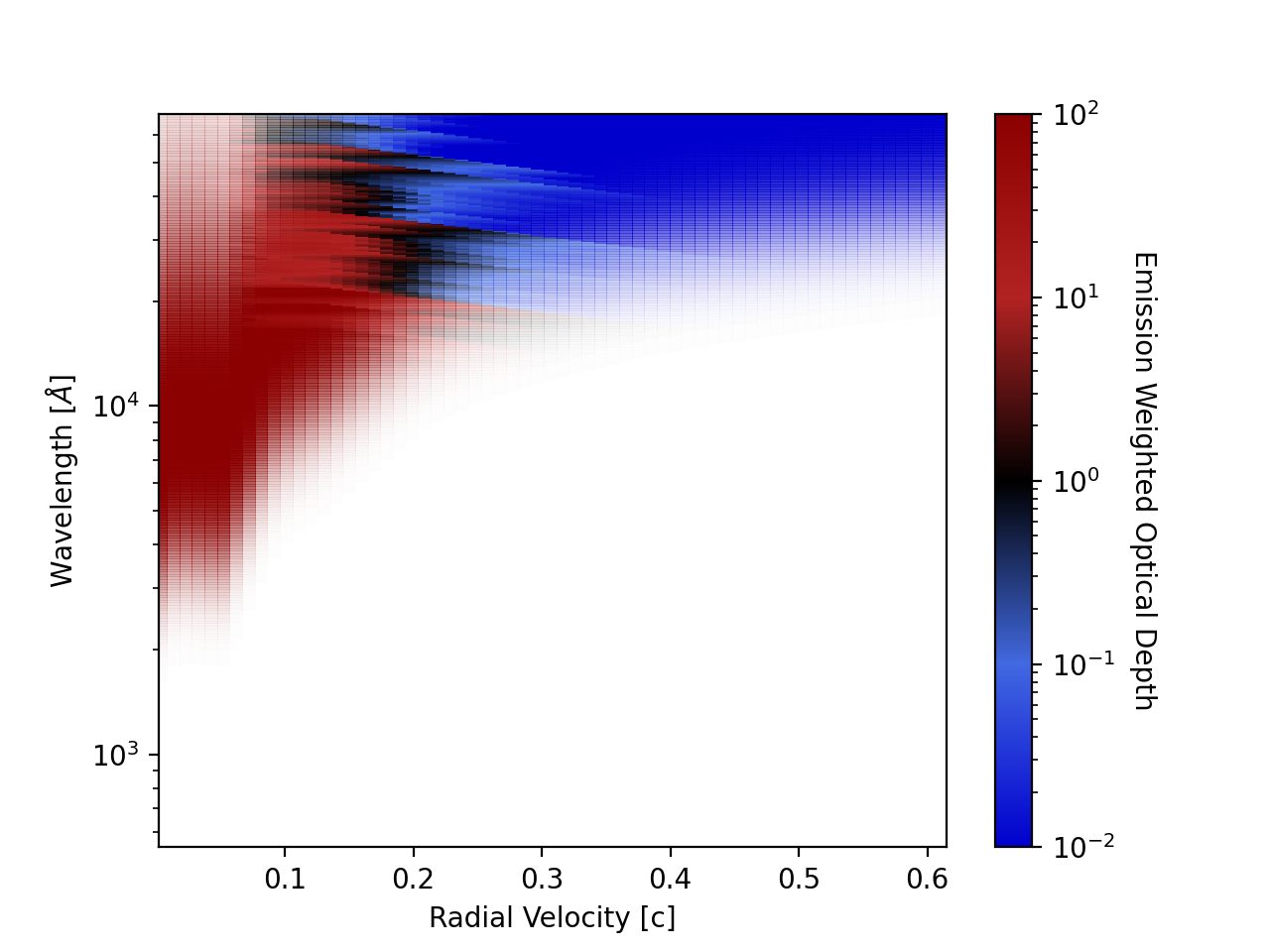}
\includegraphics[width=3.5in]{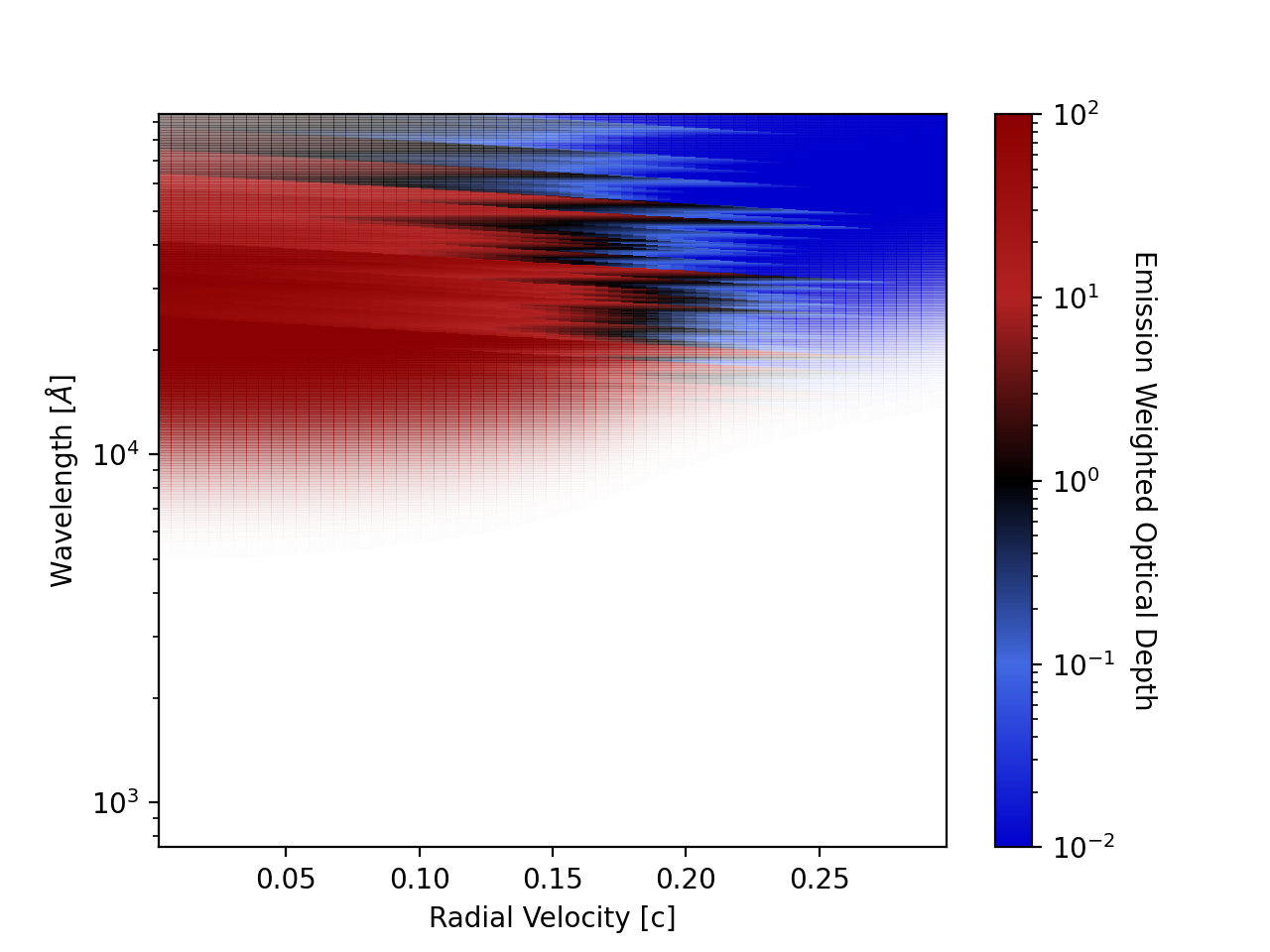}
\includegraphics[width=3.5in]{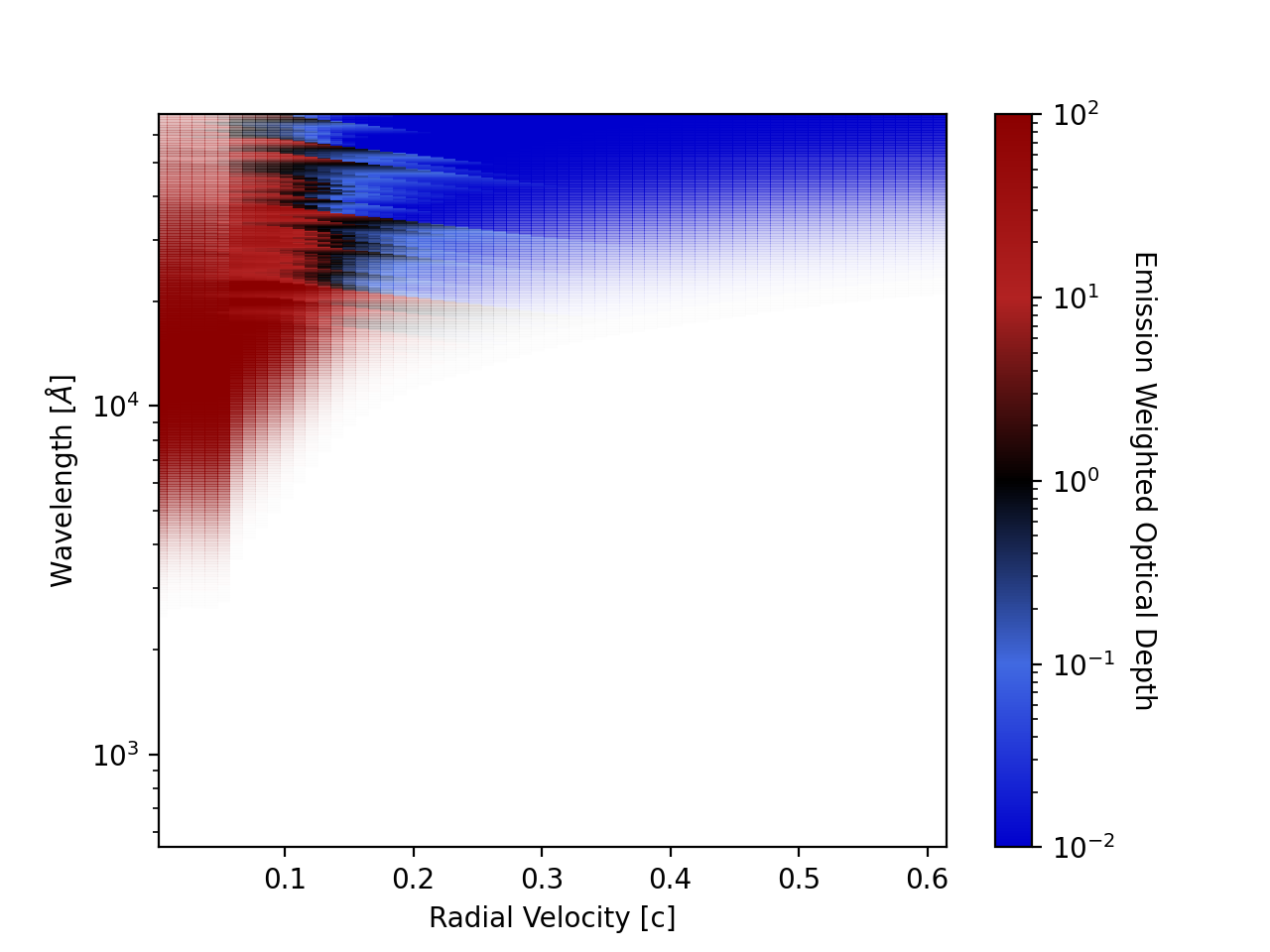}
    \caption{Emission weighted opacity for the same two models (W1 - left, M1 - right) in figure~\ref{fig:suite} at 3 different times:  2 (top), 5 (middle) and 12.5 (bottom) days.  With time, the emitted spectra move redward and the ejecta becomes increasingly optically thin.  The white regions correspond to regions where the emission is negligible.  In all cases, the optical depth moves inward (in velocity/radius space) more quickly at higher wavelengths.  Depending on the model, 2-5\,d marks the transition when the emission is entirely above 10,000\AA.}
    \label{fig:suitetime}
\end{figure}

Figure~\ref{fig:suitetime} shows a time series of these combined plots at 2, 5, and 12.5\,d for the same two sample models from Figure~\ref{fig:suite}.  Although there is material emitting in the optical and UV in the M1 model, this material is so deep in the ejecta that it is trapped.  For this M1 model, even at 12.5\,d, the radiation remains trapped at all wavelengths in the innermost region.  For our W1 model at 12.5\,d, the entire ejecta is optically thin at the highest wavelength ranges.  

\begin{figure}
    \centering
    \includegraphics[width=5in]{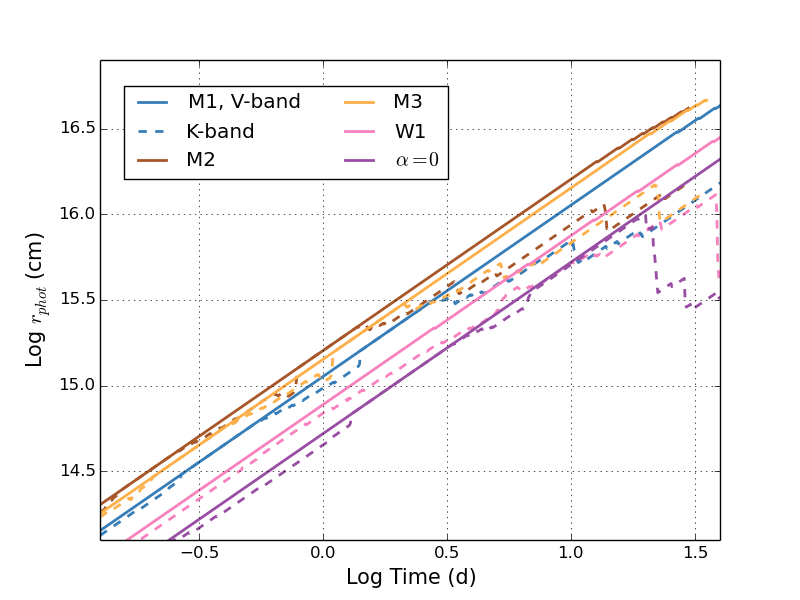}
    \caption{Photospheric radius for the V- and K-bands as a function of time.  The V-band, K-band photospheres are given by the optical depth at 5,450~\AA\  and 22,000~\AA\  respectively.  The higher opacities in the UV and visible bands (as shown in the V-band) indicate that the photosphere is further out than the K-band photosphere.  The sudden jumps in the optical depth are caused by opacity shifts at the specific wavelengths chosen for the photosphere.}
    \label{fig:rphot}
\end{figure}

\begin{figure}
    \centering
    \includegraphics[width=5in]{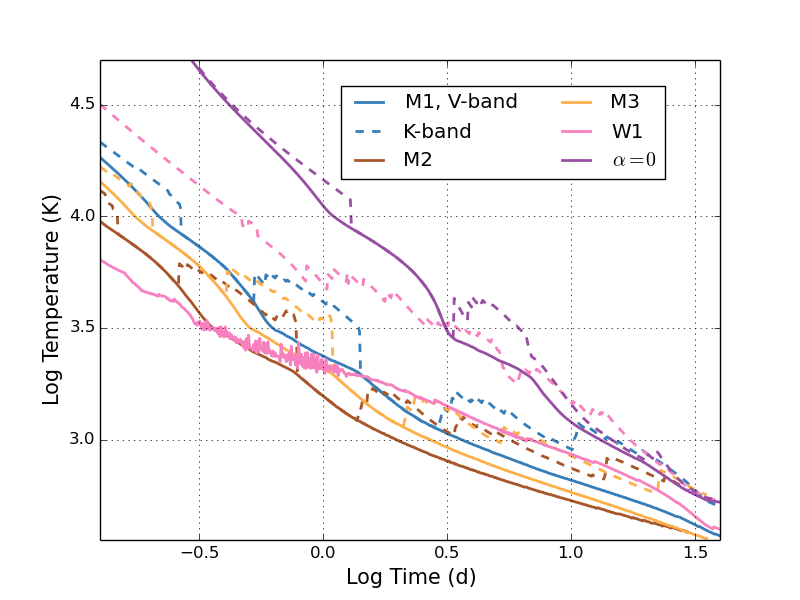}
    \caption{Temperature at the V- and K-band photospheres for the subset of the models shown in Figure~\ref{fig:rphot}.  The rapid expansion of the ejecta causes it to quickly cool.  In many of the models, the temperature at the K-band photosphere drops below 3,000~K in less than a day.  For a slightly smaller subset of V-band photospheres, this temperature also drops below 3,000~K.}
    \label{fig:temp}
\end{figure}

\begin{figure}
    \centering
    \includegraphics[width=5in]{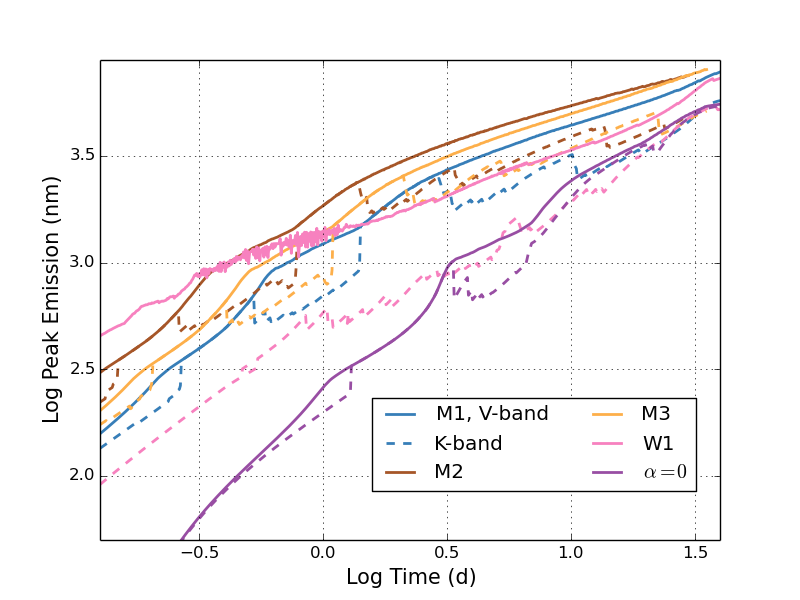}
    \caption{Peak emission wavelength at the V- and K-band photospheres for the subset of the models shown in  Figure~\ref{fig:rphot}.  For many of the models, the peak emission rises above 1,000~nm (10,000~\AA) at 1\,d.}
    \label{fig:emit}
\end{figure}

Figures~\ref{fig:suite} and \ref{fig:suitetime} contain a broad set of information, showing photospheres and emission as a function of both wavelength and position.  To better understand how these structures map to time-dependent light-curves, we can focus on the evolution of the photosphere in specific bands.  Figure~\ref{fig:rphot} shows the position of the V-band and K-band photospheres as a function of time.  The heavy elements in neutron star merger ejecta lead to high opacities in the UV and optical bands, causing the V- and UV-band photospheres (we only show the V-band) to be further out than the K-band.  At the V-band photosphere, the temperature quickly drops sufficiently such that the peak emission wavelength increases to progressively redder bands (see Figures~\ref{fig:temp},~\ref{fig:emit}).  It is this evolution that leads to the K-band dominating the emission at late times.

These plots provide clues into the double-peaked structure of the bolometric light-curve.  This double-peaked structure is dominated by the infra-red emission as is seen in our K-band images.  The sharp drop in the emission for our models occurs because the lanthanide opacity increases dramatically as the temperature drops from 10,000 to 5,000K.  This moves out the photosphere, reducing the temperature at the photosphere, causing a drop in the emission.

The figures also demonstrate one of the difficulties in determining the exact heavy-element yields in kilonova from band measurements alone.  For many of our models, the temperature of the photosphere (for either the V- or K-bands) after 1~day is so low that the emission peaks in the K-band (Fig.~\ref{fig:emit}).  This means that the emission would peak in the K-band even if there were no heavy elements trapping the optical emission at higher radii.  That is, for some models, the late-time infra-red emission could be due to cooling of the ejecta and not just from a lanthanide ``curtain".  This explains why the emission from GW170817 could be fit by a broad range of models with very different heavy r-process mass fractions~\citep{2018ApJ...855...99C}.  Although late-time IR emission is suggestive of lanthanide production, it is not a proof that a large amount of r-process elements were produced in the ejecta.

Finally, we can use these figures to help us understand the different behavior of the models.  The distribution mass as a function of velocity sets the evolution of the photosphere.  For example, in our disk models (M series), the photosphere remains high out to late times.  This means that the temperature is lower and the UV- and V-band lightcurves evolve quickly.  Our power-law ($\alpha$) models have the smallest photospheres, probing hotter regions that allow the UV- and V-bands to remain bright longer.  The K-band is less sensitive to the velocity distribution because the photospheric differences tend to be less at higher wavelengths. 

We have not included all of the effects of the velocity distribution in these models.  For instance, we use a simple density and time-dependent formula for the energy deposition from electrons and alpha particles, but these different velocity structures will change the fraction of energy deposited in different regions.  

\subsection{Dependence on Other Properties}

To compare to light-curve variations produced by other properties of the ejecta, we have included in our study a set of models where we vary both the ejecta mass and energy, and another set where we vary the composition.  In this section, we study the dependence on these two properties.

\begin{figure}[ht]
\includegraphics[width=3.5in]{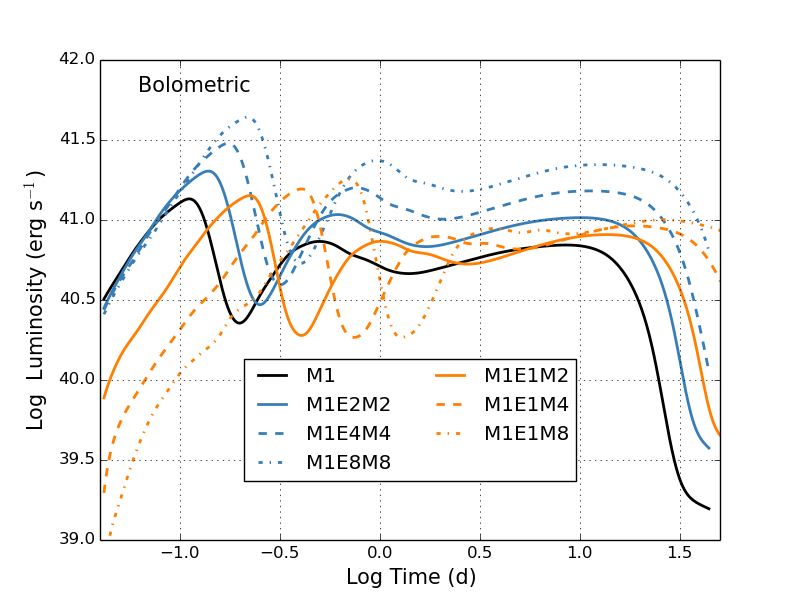}
\includegraphics[width=3.5in]{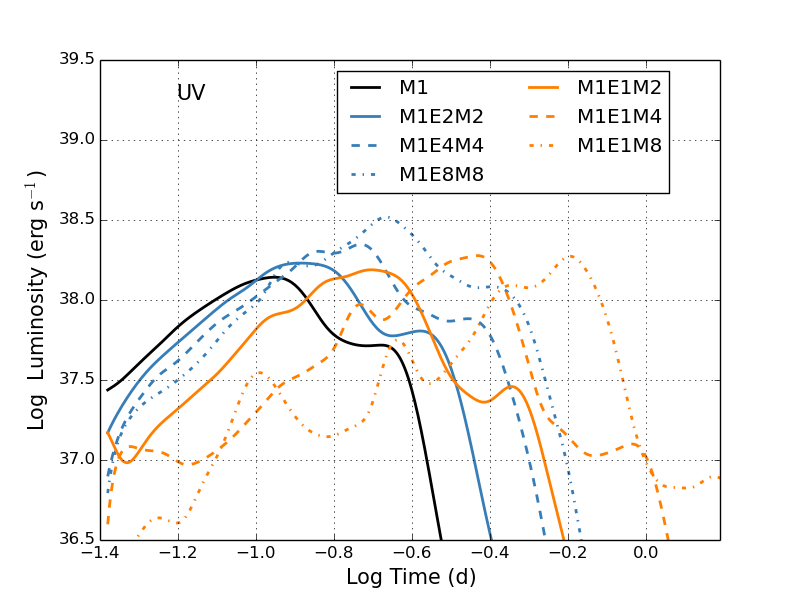}
\includegraphics[width=3.5in]{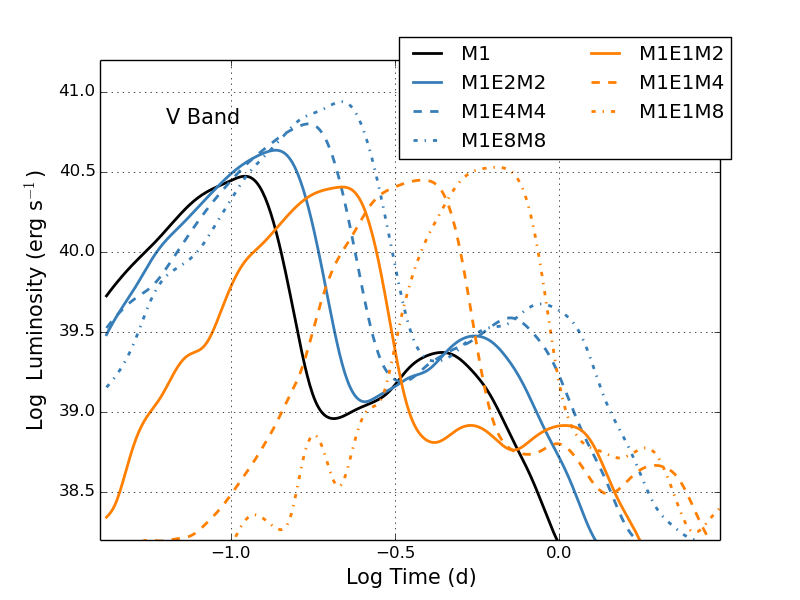}
\includegraphics[width=3.5in]{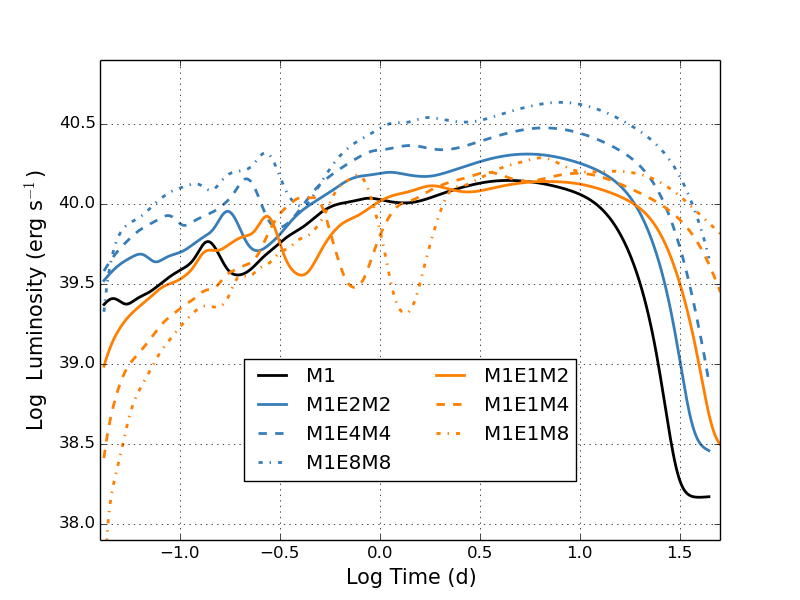}
    \caption{Evolution of UV, V-band and K-band luminosities with respect to the ejecta mass and energy.  We use the M1 model as the standard and then increase the mass by 2, 4 and 8 times (M2,M4,M8).  We include two sets of models:  one where the energy increases with mass (${\rm Energy \propto Mass}$) with E2, E4, E8 corresponding to increasing the energy by 2, 4, and 8 times.  This corresponds to holding the velocity constant.  In another suite, we hold the energy fixed and the velocity decreases with ejecta mass to the 1/2 power.}
    \label{fig:mass}
\end{figure}

Figure~\ref{fig:mass} shows the dependence of our bolometric, UV-, V- and K-band light-curves on the mass of the ejecta.  Using our M1 disk velocity distribution, we increase the mass by 2, 4 and 8 times (E2M2, E4M4 and E8M8 have masses 0.2, 0.4, 0.8 $M_\odot$ but keep the same velocities).  For our first suite of models, we increase the explosion energy proportional to the mass so that the ejecta velocity remains constant.  From Figure~\ref{fig:mass} we see many expected trends with ejecta mass ($M_{\rm ejecta}$).  For example, the higher the mass, the longer it takes the emission to peak (the increased optical depth means that the ejecta layers become optically thin at later times).  The delay in the peak emission scales as $M_{\rm ejecta}^{0.2-0.5}$, depending on the band.  The larger energy source (more mass means more radioactive material) also produces brighter peak emission.  Although the total energy released scales roughly as the ejecta mass, the longer timescales cause the peak luminosity to scale at a lower power, closer to ($M_{\rm ejecta}^{0.5}$).  The exact value depends on the waveband and the model.  This scaling of the peak luminosity and timescale matches fairly well the mass scaling from Equations 27 and 28 of \cite{2018MNRAS.478.3298W}.  But the fact that the variations depend on the wavelength range (band pass) partially explains the range of correlations predicted by different groups.  

We have also run a set of models where we increase the ejecta mass but hold the energy constant (E1M2, E1M4, E1M8 corresponding to have masses 0.02, 0.04, 0.08 $M_\odot$ but the same energy).  In this case, the velocity decreases with increasing ejecta mass (velocity $\propto M_{\rm ejecta}^{-0.5}$).  For most of the light-curve bands in our models, the peak time increases with the inverse of the velocity.  This effect is stronger than the $v^{-0.6}$ dependence predicted by \cite{2018MNRAS.478.3298W}, but expected in 1-dimensional scenarios where the radiation is trapped with the flow and only the outer material is radiating.  The fact that the K-band is sensitive to the mass and velocity, but not so much to the velocity distribution, means that combined K- and UV- or V-band light-curves will allow us to distinguish between these two effects insofar as we can disentangle both from other model systematics.

\begin{figure}[ht]
\includegraphics[width=3.5in]{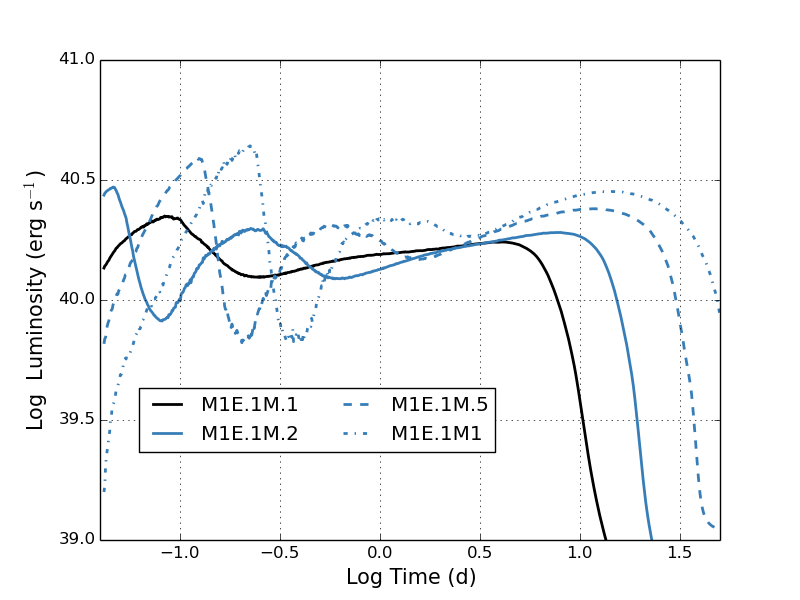}
\includegraphics[width=3.5in]{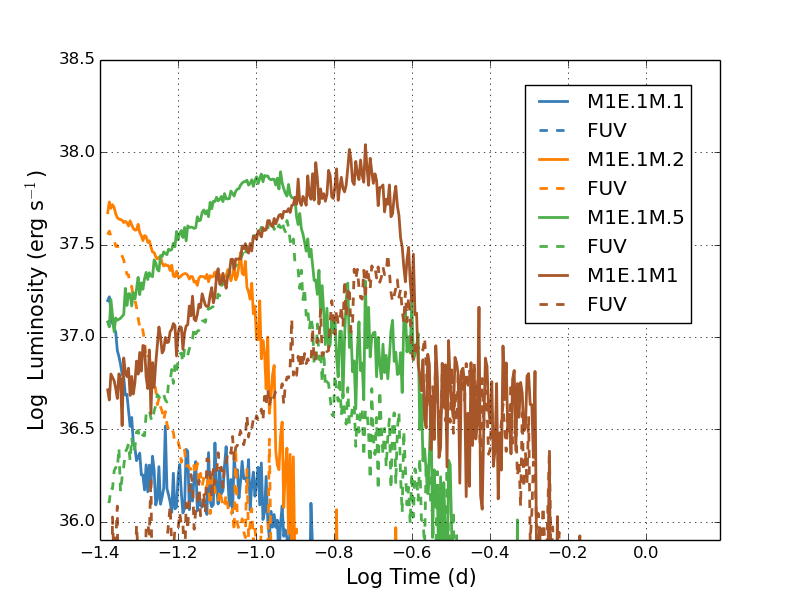}
\includegraphics[width=3.5in]{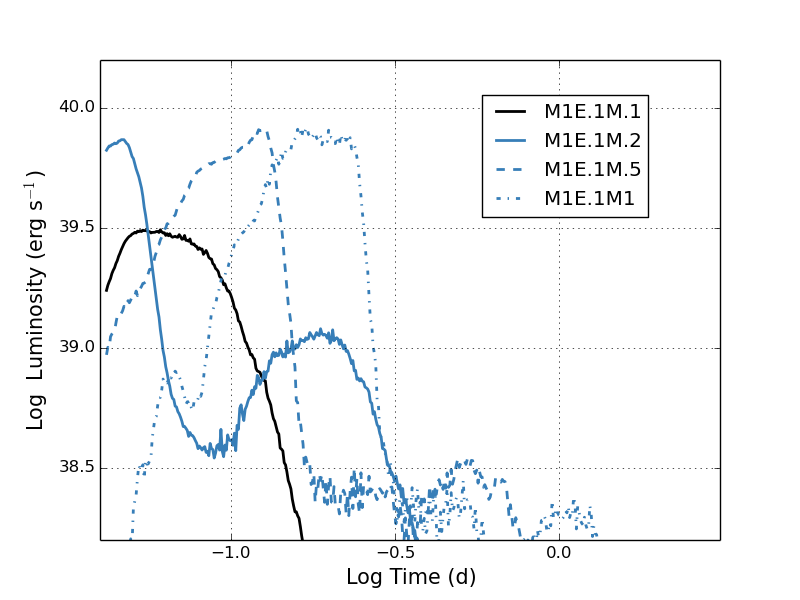}
\includegraphics[width=3.5in]{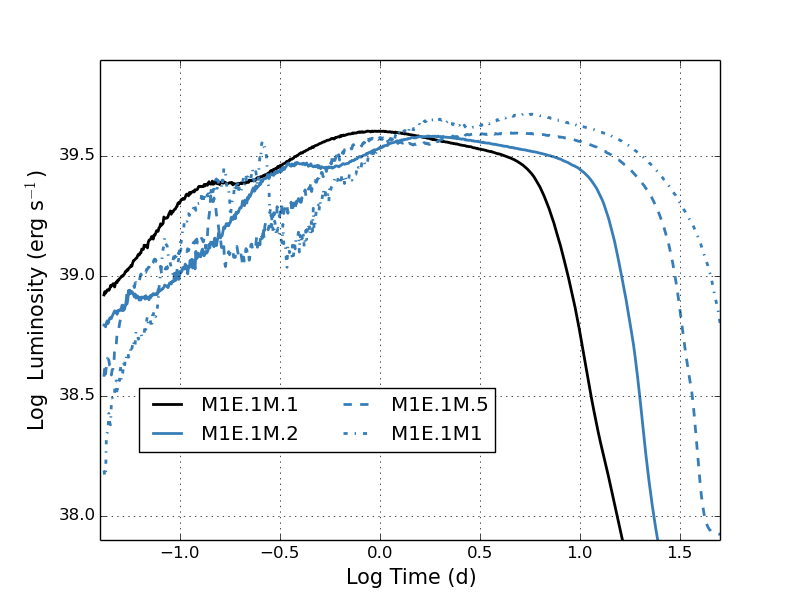}
    \caption{Evolution of UV, V-band and K-band luminosities with respect to the ejecta mass and energy.  We use the M1 model as the standard, lowering the energy by a factor of 10 and using 4 masses that are 0.01, 0.02 0.05 and 0.1M$_\odot$.  For the UV plot, we also include a series of far UV signals denoted by dashed lines (based on the wavelength sensitivity of UVEX). Since the energy is held constant in these models, increasing the mass also leads to lower velocities.  Lower velocities lead to slower time evolution.  The V-band and ultraviolet emission is most sensitive to the structure of the outer layers.  The K-band is more sensitive to the total mass.}
    \label{fig:mass2}
\end{figure}

These are high ejecta masses.  Figure~\ref{fig:mass2} shows the ejecta mass for our M1 disk velocity distribution using masses of 0.01, 0.02, 0.05 and 0.1 $M_\odot$ all with an explosion energy of $4 \times 10^{50}\,{\rm erg}$.  The trends, especially in the bolometric and K-band luminosities follow the same trends discussed above.  UV and, to a lesser extent, V-bands, whose early-time peak luminosities depend on the outermost ejecta, are less sensitive to the total mass and more dependent on the outer structure.  Because we keep the critical velocity for the power-law shift and the we keep the total energy fixed, the trends in the UV and V-bands are more complex than the simple models predict.

\begin{figure}[ht]
\includegraphics[width=3.5in]{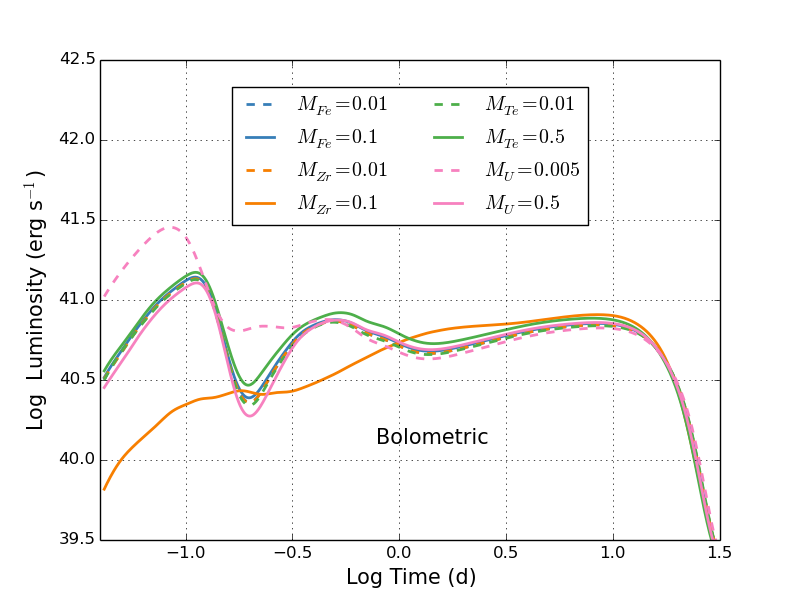}
\includegraphics[width=3.5in]{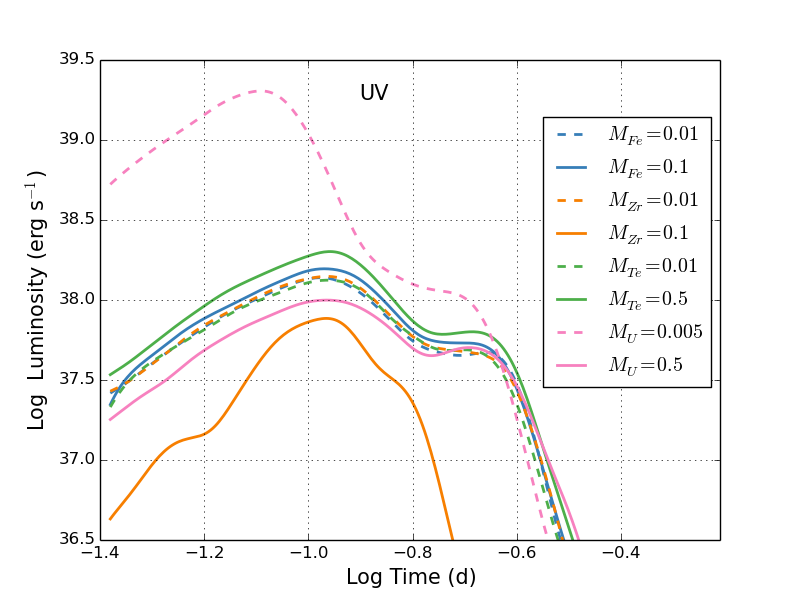}
\includegraphics[width=3.5in]{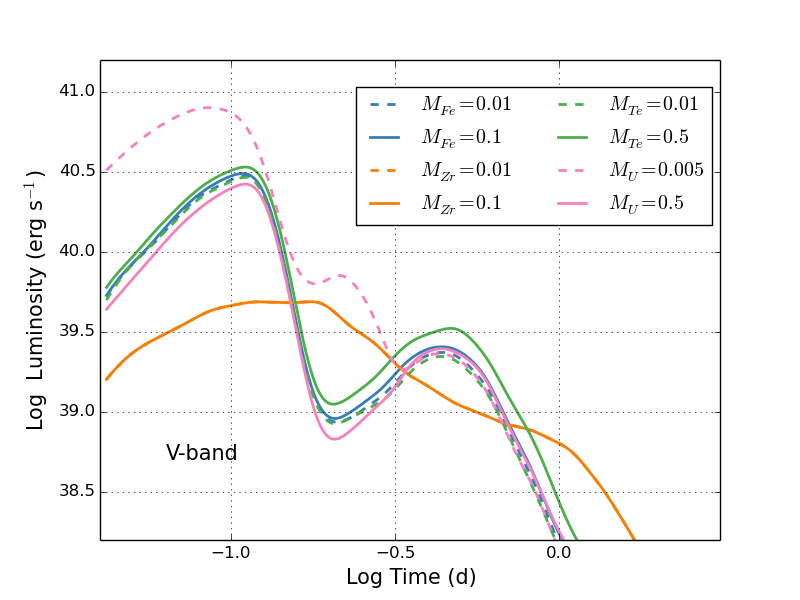}
\includegraphics[width=3.5in]{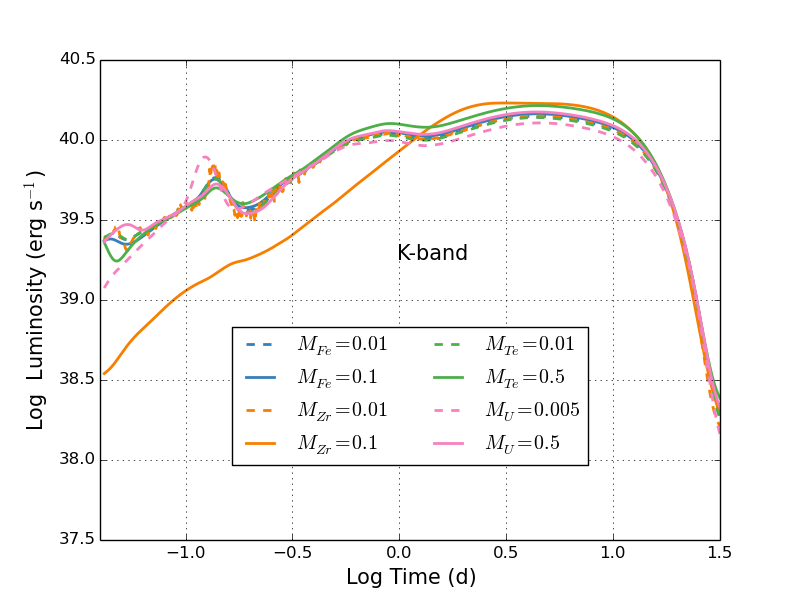}
    \caption{Evolution of UV, V-band and K-band luminosities with respect to the ejecta mass and energy.  We use the M1 model set as the standard and then vary individual elements (raising and lowering each with respect to our standard).  When we modify one element, we scale up or down all other mass fractions to keep a constant mass fraction.  The most dramatic effect is the uranium mass fraction and it is strongest in the UV.}
    \label{fig:abun}
\end{figure}

Opacities depend on the density and temperature (which depends on the velocity distribution), but they also depend on the composition.  To understand these effects, we varied the abundances in our models by raising and lowering the mass fractions of different elements ranging from iron to uranium (e.g. M1Fe0.01).  We then scale the remaining elements to keep a constant mass fraction.  In this study, we only vary the opacities, not the energy deposition.  For the most part, the exact element abundance has a small effect on the light-curves (Figure~\ref{fig:abun}), confirming the results from ~\cite{2020ApJ...899...24E}.  But varying the uranium abundance can make a large difference in the light-curves, especially in the UV.  In the UV, lowering the uranium abundance can raise the peak luminosity by over an order of magnitude.  This effect is present, but less extreme, in the V-band.  Altering the Zr abundance can alter not just the UV- and V-bands, but also the K-band emission, confirming past results~\citep{2020ApJ...899...24E}.

\section{Spectra}

Spectra provide additional opportunities to differentiate the velocity distribution from other properties of the ejecta.  In this section, we present time-dependent broadband spectra from a subset of our models.  Because the UV bands are such strong discriminants, we focus our study on the utility of UV spectra.  Given the portfoloio of proposed ultraviolet missions including the phase A-selected UVEX mission and the SIBEX shock interaction proposal, ultraviolet spectra may be achievable in the near future.

\begin{figure}[ht]
\includegraphics[width=3.5in]{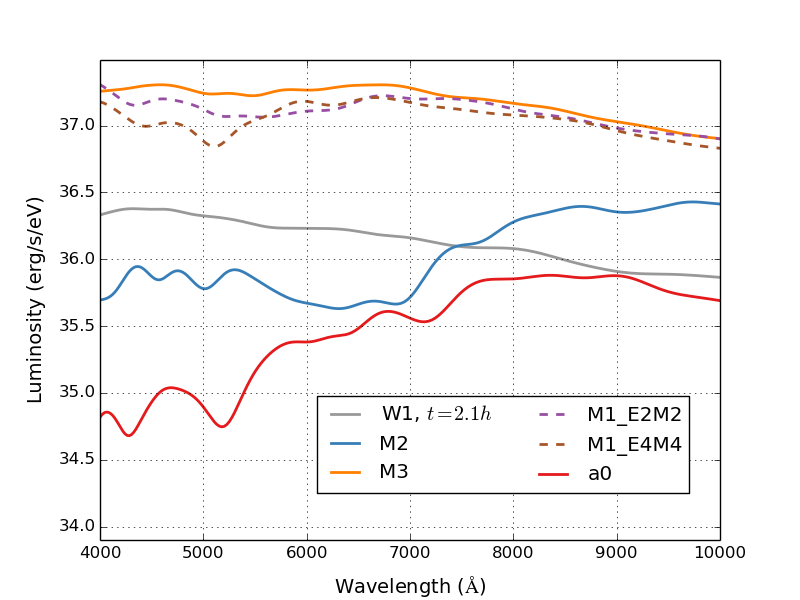}
\includegraphics[width=3.5in]{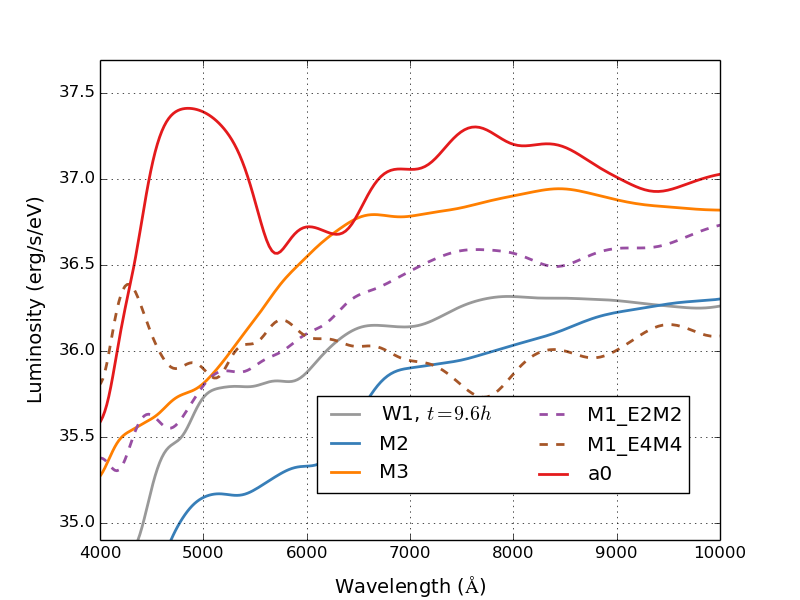}
\includegraphics[width=3.5in]{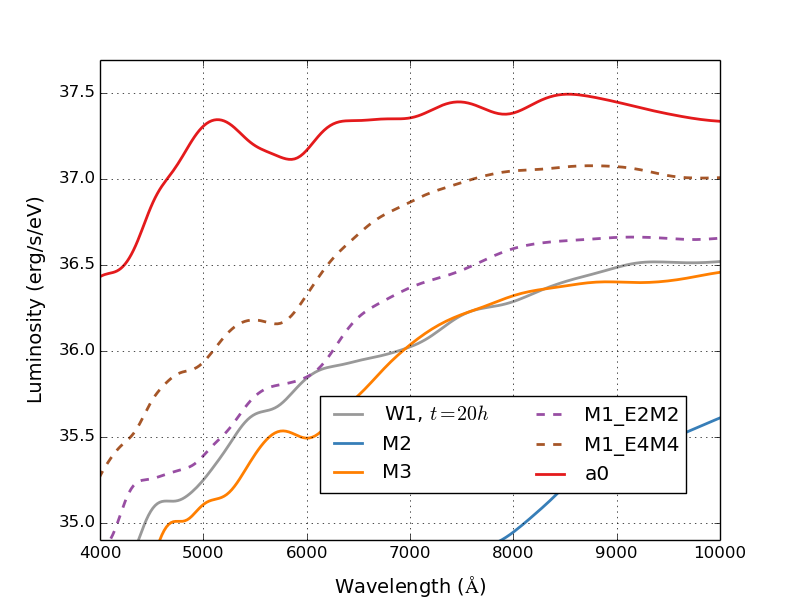}
\includegraphics[width=3.5in]{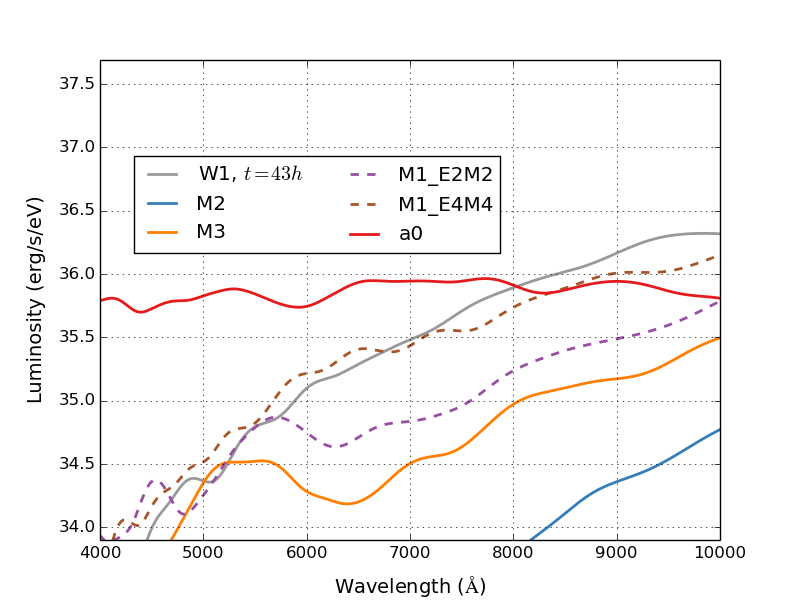}
    \caption{Spectra versus wavelength at four times (2.1, 9.6, 20, and 43h) for a range of models:  w1, M3, M4, $\alpha=0$ from Figure~\ref{fig:morph} and E2M2, E4M4 from Figure~\ref{fig:mass} (see Table~\ref{tab:models} for a description).  Very few strong features appear in the spectra (although note the broad feature in the $\alpha=0$ model at 9.6h) and most have very similar shapes.}
    \label{fig:specfull}
\end{figure}

Figure~\ref{fig:specfull} shows spectra from 4,000--10,000~\AA\ for six of our models including both standard, power-law, and a series of disk-wind guided velocity distributions.  We include two models with an increased mass.  As the ejecta expands and the temperature at the photosphere cools, the spectra become increasingly red, with the bulk of the emission occurring at longer wavelengths.  The forest of lines coupled with high velocities producing Doppler broadening produce fairly smooth structures, but some broad features do appear, especially in our power-law distributions.  The nature of the broad features depends upon the velocity distribution, meaning that there is not a one-to-one correspondence between the features and the composition.  In addition, many of the spectral features of one model can be mimicked by the spectral features of another at a slightly different time in the evolution and can also be reproduced by different ejecta masses.  In turn, it will be hard to determine the ejecta masses without constraining the velocity distribution.

\begin{figure}[ht]
\includegraphics[width=3.5in]{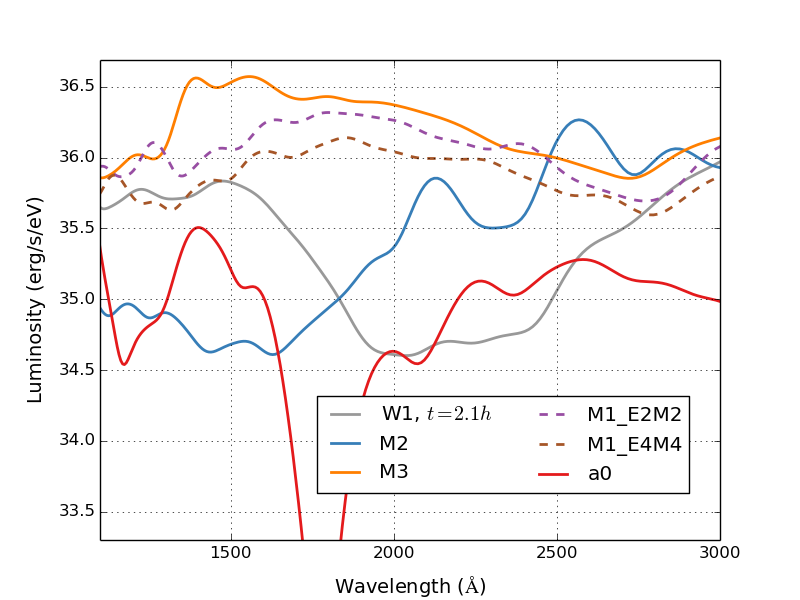}
\includegraphics[width=3.5in]{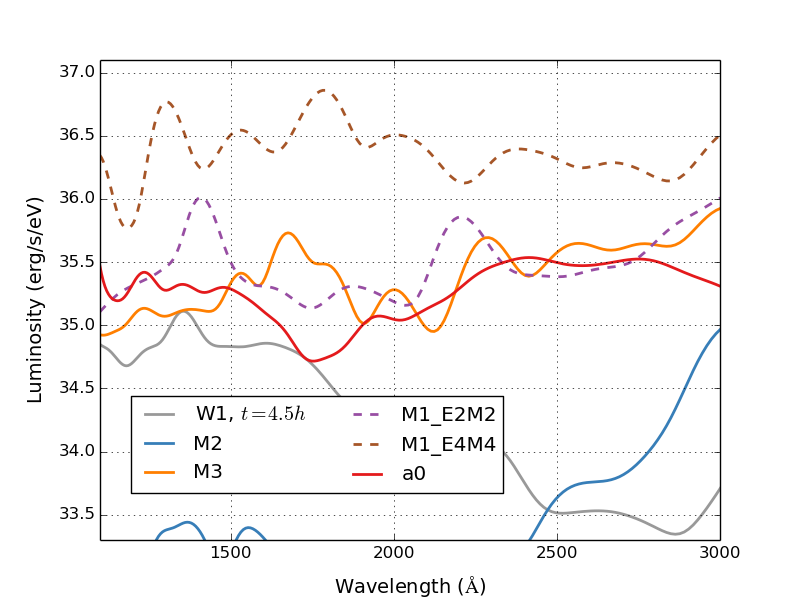}
    \caption{UV spectra (1000--3000~\AA) at 2.1h and 4.5h for the subset of our models described in Figure~\ref{fig:specfull}. As with the UV light-curves, the spectra of the varied velocity-distribution models evolve very differently.}
    \label{fig:specuvel}
\end{figure}

The UV emission has more features, but is difficult to observe except at early times.  Figure~\ref{fig:specuvel} shows the spectra of our different models at 2.1 and 4.5h.  Bear in mind that our opacities are calculated using only the first 4 ion stages and their accuracy diminishes above 20,000K and, especially for our a0 ($\alpha=0$) model, the photosphere temperature exceeds this value for the first 15 hours.  The lack of higher ion stages will affect the spectra for this model at early times.  However, for most of our other models, the photosphere is below 20,000K within 2 hours of the explosion.  The interior may be hotter than this, but because the radiation is fairly diffusive in the interior, missing the higher ion stages will not have a large effect on the emission.  The evolution of the different velocity distributions is very different and, if multiple measurements can be observed at these early times, uncertainties in the velocity distribution can be disentangled from the ejecta mass.  Such early observations could help us better use the optical emission to determine other ejecta properties.  

\begin{figure}[ht]
\includegraphics[width=3.5in]{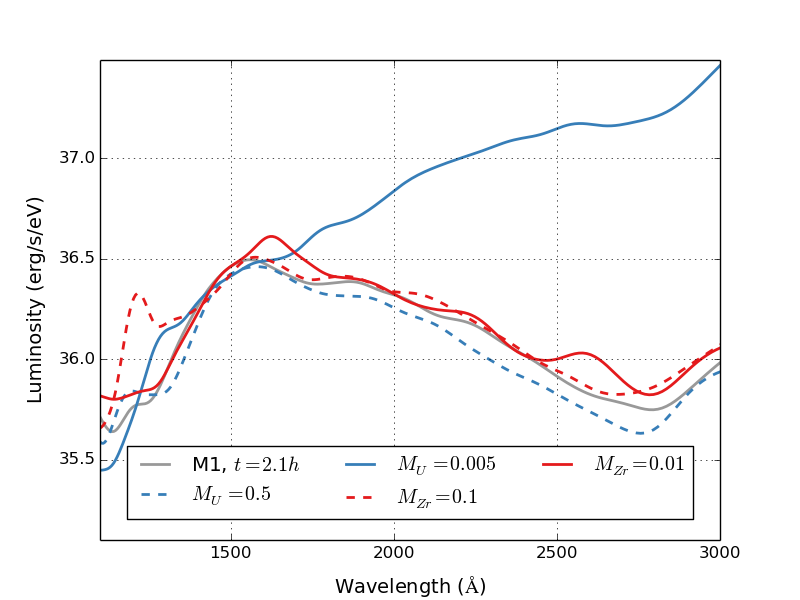}
\includegraphics[width=3.5in]{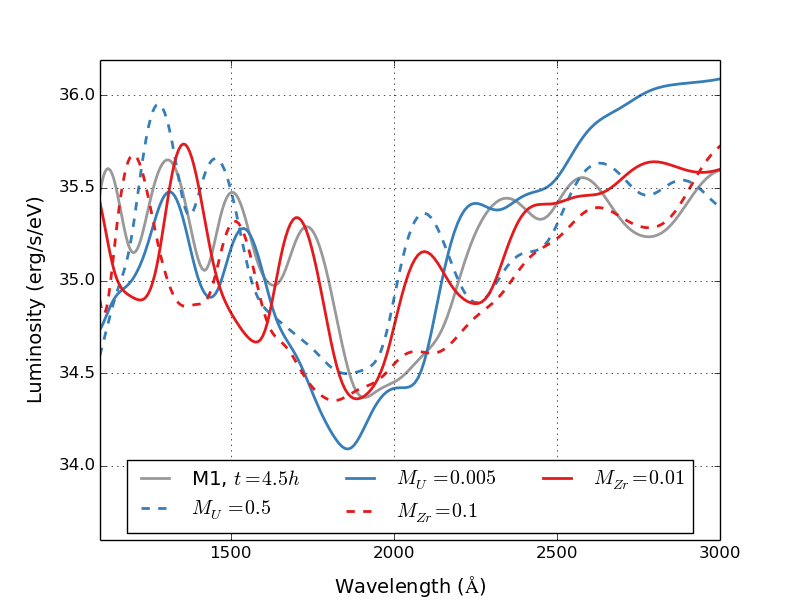}
    \caption{UV spectra (1000--3000~\AA) at 2.1h and 4.5h for models varying the composition.  Lowering the uranium abundance allows radiation at higher wavelengths to escape.  Line features in zirconium are indicative to of the importance of 4th and 5th row elements in the wind ejecta.}
    \label{fig:specuvelabun}
\end{figure}

The broad line features may also provide some clues about the composition.  Figure~\ref{fig:specuvelabun} shows the UV spectra at these early times.  As with the UV light-curves, reducing the uranium mass fraction allows for much more emission at higher UV wavelengths.  Although specific line features may be difficult to detect, the emission at these wavelengths will help determine the produce of this element.  The zirconium abundance produces much more varied line features and the UV features might be able to constrain the fractions of 4th and 5th row elements.

Spectra are ideally suited to help differentiate models.  But it is important to realize that it is often difficult to disentangle the broad features produced by the blending of lines, especially when uncertain features, such as the velocity distribution, exist.  The spectra are incredibly sensitive to the conditions of the photosphere.  Features in a given model can shift dramatically depending on these conditions.  The structure of $a0$ model produces a distinct photosphere.  This produces strong features in the spectra.  But the specific features depend sensitively on the initial conditions.  A wide-ranging investigation of input physics and systematics will be needed to make sufficiently robust predictions about the relationship between the emitted light and ejected material in kilonovae. 

\section{Summary}
\label{sec:dependence}

The electromagnetic signal from the non-relativistic ejecta from merging neutron star binaries provides a means to determine both the total mass and, in particular, the mass of r-process material ejected in these mergers.  However, the observed light-curves are sensitive to not just the ejecta mass, but to their energy, morphology, composition and, as we have shown in this paper, the velocity distribution of the ejecta.  Nevertheless, because the effect that each of these ejecta properties has on the light-curves varies on the observed energy band, broadband coverage of a kilonova event has the potential to differentiate the effects, allowing astronomers to produce more accurate ejecta yields.  In this summary, we discuss the importance of this broadband coverage.

The UV- and V-bands are most sensitive to the velocity distribution because they are the most sensitive to the outermost photospheric evolution (see Figure~\ref{fig:morph}).  The K-band is less sensitive to this distribution, but it still can vary by a factor of three at 5-10~days past peak.  From Figure~\ref{fig:mass}, we see that varying the mass by a factor of 4 varies (compare model M1E2M2 to M1E8M8) the K-band luminosity factor of 2.5. With these models as a guide, the variation in K-band luminosity caused by the velocity distribution corresponds to variations produced by varying the mass by at least a factor of 4.  If we include power-law distributions (often used in the literature), this variation can be a factor of 4.  But because the UV- and V-bands are so sensitive to the velocity distribution, we can use them to constrain the velocity distribution, allowing K-band observations to be used to more accurately study the ejecta mass. 

Although broad composition changes (e.g. existence of lanthanides) can have strong effects on the light-curves, it is harder to identify individual composition variations from the light-curves.  The luminosity after a few days (primarily in the K-band) was fairly insensitive to the composition variations used in this paper.  More detailed composition studies have already been done~\citep[e.g.][]{2020ApJ...899...24E,2022MNRAS.tmp.2583F,2022ApJ...939....8D} and we point the reader to these studies.  Because the UV emission is particularly sensitive to the uranium abundance, if we can use other bands to constrain the the ejecta mass and velocity distribution, we may be able to use the UV to constrain the amount of uranium in the ejecta.

Even though the density of lines and large velocity gradients make it difficult to identify specific line features in the data~\citep[see, for example][]{2022ApJ...939....8D}, broad spectral features have the potential to place even further constraints on the ejecta properties.  Coupling detailed spectra with light-curve models provides a powerful means to disentangle this physics.  

By studying the relative photospheres at different wavelengths, we find that, depending on the velocity distribution, the temperature at the photosphere (even in the K-band) can be quite low after a few days and, for some models, the strong infra-red signal in the emission is not evidence of a lanthanide ``curtain", but more an indication of how quickly this fast-moving ejecta cools.  Interpreting the existence of lanthanides based on late-time infra-red observations requires an understanding of the ejecta velocity distribution.

This paper primarily focused on the uncertainties in the velocity distributions for lanthanide-rich ejecta.  Another paper was recently submitted studying ejecta properties using the Tardis code~\cite{2023arXiv231015608T}.  This paper focused on the evolution of the spectra using a formula for the evolution of the photosphere temperature.  These calculations are ideally suited for showing the spectral dependencies of the velocity distribution and confirms the importance of the velocity distribution on the observed emission.  Both our study and this new one demonstrate the importance of this velocity distribution.

Other studies have focused on other ejecta properties.  To truly determine how to disentangle all of these effects, a broader grid of simulations that studies all of these effects is important.  In addition, physical uncertainties including atomic physics (including the fact that we only included the first 4 ion stages), nuclear physics, out-of-equilibrium and energy deposition effects can also drastically alter the light-curve models.  Much more work needs to be done to identify the key light-curve and spectral features of each effect so that we can disentangle each of them.  Nevertheless, if we can correctly identify and understand these effects, kilonova observations have the potential to constrain not only the ejecta masses and compositions (which are important for understanding r-process production), but also the physics that is responsible for producing this ejecta (e.g. nuclear physics).

\begin{acknowledgements}

The work by CLF, JMM, SD, and CJF was supported by the US Department of Energy through the Los Alamos National Laboratory (LANL). JMM was additionally supported by the Laboratory Directed Research and Development program of LANL through the Centers for Earth and Space Science (CSES) and Nonlinear Studies (CNLS) under project numbers 20210528CR, 20220545CR-CNL, and 20220564ECR. SD acknowledges support from the Director's Postdoctoral Fellowship at LANL funded by the Laboratory Directed Research and Development Program project number 20200687PRD2.
This research also used resources provided by LANL through the institutional computing program.
Los Alamos National Laboratory is operated by Triad National Security, LLC, for the National Nuclear Security Administration of U.S.\ Department of Energy (Contract No.\ 89233218CNA000001).  Part of this work was done at the Aspen Center for Physics, which is supported by National Science Foundation grant PHY-1607611.

\end{acknowledgements}

\bibliography{refs}{}
\bibliographystyle{aasjournal}



\end{document}